\documentclass[12pt]{article}
\usepackage{amsfonts,amsmath,amssymb,epsf}
\usepackage{amsmath,braket}
\usepackage{comment}

\usepackage{graphicx,hyperref,color}
\hypersetup{
    unicode=false,          
    pdftoolbar=true,        
    pdfmenubar=true,        
    linktocpage=true,       
    pdffitwindow=false,     
    pdfstartview={FitH},    
    pdfnewwindow=true,      
    colorlinks=true,       
    linkcolor=blue,          
    citecolor=blue,        
    filecolor=blue,      
    urlcolor=blue           
}
\numberwithin{equation}{section}									

\def\d#1{\,{\rm d}#1}

\newcommand{\sgn}{\textrm{sgn}}
\newcommand{\la}{\langle}
\newcommand{\ra}{\rangle}
\newcommand{\Tr}{\textrm{Tr}}
\newcommand{\Pf}{\textrm{Pf}}
\newcommand{\Dig}{\textrm{Dig}}
\newcommand{\IR}{\textrm{IR}}

\newcommand{\WH}{\scriptsize\mbox{WH}}
\newcommand{\HW}{\scriptsize\mbox{HW}}
\newcommand{\onshell}{\scriptsize\mbox{on-shell}}
\newcommand{\oneloop}{\scriptsize\mbox{one-loop}}
\let\a=\alpha \let\b=\beta \let\c=\chi \let\d=\delta      \let\l=\lambda \let\m=\mu \let\n=\nu
\let\o=\omega   \let\s=\sigma \let\t=\tau \let\th=\theta  \let\vp=\varphi   
 \let\D=\Delta \let\G=\Gamma    \let\S=\Sigma    
\def\nn{\nonumber}
\def\inf{\infty}

\def\pa{\partial}

\def\wtd{\widetilde}

\begin{document}

\begin{titlepage}
\thispagestyle{empty}

\vspace*{-1cm}
\begin{flushright}
RIKEN-iTHEMS-Report-21
\\
YITP-21-124
\\
\end{flushright}

\bigskip

\begin{center}
\noindent{{\Large \textbf{Factorizing Wormholes\\ \vspace{5mm}
in a Partially Disorder-A\hspace{-0.6mm}veraged SYK Model}}}\\
\vspace{2cm}

\quad Kanato Goto$^a$, \, Kenta Suzuki$^b$ \ and \ Tomonori Ugajin$^{b}$
\vspace{1cm}\\

{\it $^a$\it  RIKEN Interdisciplinary Theoretical and Mathematical Sciences\\ (iTHEMS), Wako, Saitama 351-0198, Japan}\\
\vspace{1mm}
{\it $^b$Center for Gravitational Physics,\\
Yukawa Institute for Theoretical Physics,
Kyoto University, \\
Kitashirakawa Oiwakecho, Sakyo-ku, Kyoto 606-8502, Japan}\\

\vspace*{2cm}
\vskip 2em
\end{center}

\begin{abstract}

In this paper, we introduce a ``partially disorder-averaged'' SYK model. 
This model has a real parameter that smoothly interpolates between the ordinary totally disorder-averaged SYK model and the totally fixed-coupling model. 
For the large $N$ effective description, in addition to the usual bi-local collective fields, we also introduce a new additional set of local collective fields.
These local fields can be understood as ``half'' of the bi-local collective fields, and in the totally fixed-coupling limit,
they represent the ``half-wormholes'' which were found in recent studies.
We find that the large $N$ saddles of these local fields vanish in the total-disorder-averaged limit,
while they develop nontrivial profiles as we gradually fix the coupling constants.
We argue that the bulk picture of these local collective fields represents a correlation between a spacetime brane and the asymptotic AdS boundary.
This illuminates how the half-wormhole saddles emerge in the SYK model with fixed couplings.

\end{abstract}

\end{titlepage}

\newpage

\tableofcontents

\section{Introduction}

Recent studies suggested a holographic dual of a bulk gravitational theory has to involve an ensemble average of theories with random couplings.
The Sachdev-Ye-Kitaev (SYK) model \cite{Sachdev:1992fk, Kitaev:2015} provides such an example of the duality. It
is a quantum mechanical many-body system with all-to-all interactions on fermionic $N$ sites ($N \gg 1$).
Its coupling constant is random with a Gaussian probability distribution.
The model provides an interesting, highly nontrivial example of the AdS/CFT duality and a potential framework for quantum black holes,
with related out-of-time-order correlators exhibiting quantum chaos
and the maximal Lyapunov exponent characteristic of black holes \cite{Kitaev:2015, Polchinski:2016xgd, Maldacena:2016hyu, Cotler:2016fpe}.
A central role is played by the emergent Schwarzian mode \cite{Maldacena:2016hyu, Jevicki:2016bwu, Bagrets:2016cdf, Jevicki:2016ito, Bagrets:2017pwq, Mertens:2017mtv, Kitaev:2017awl}, which dominates the dynamics of the model in the low-temperature limit.
The dual of the Schwarzian mode in the gravity theory is well understood as the boundary graviton of the Jackiw-Teitelboim (JT) gravity \cite{Teitelboim:1983ux, Jackiw:1984je}
in the context of nearly AdS$_2$/CFT$_1$ correspondence \cite{Almheiri:2014cka, Jensen:2016pah, Maldacena:2016upp, Engelsoy:2016xyb}.

The relationship between a boundary ensemble average and a bulk gravity theory
becomes more evident in the recent calculation of the JT gravity partition function at a non-perturbative level, defined by summing over all higher genus topologies. For instance, it was shown that the non-perturbative JT gravity partition coincides with a random Hermitian matrix integral \cite{Saad:2019lba}.
In this correspondence, the central role is played by Euclidean wormholes in the gravitational path integrals. For instance, in the presence of multiple conformal boundaries, the natural rule to define semi-classical gravitational path integral is to include all Euclidean manifolds consistent with the boundary condition. This includes Euclidean wormhole geometries where several conformal boundaries get connected, resulting in the random matrix integral, as mentioned above. It was also shown that  even in a very simple topological model of gravity, once we include all possible topologically distinct manifolds respecting the boundary condition, its boundary  naturally realizes an ensemble average \cite{Marolf:2020xie}. Its generalization to supersymmetric cases can be found in \cite{Balasubramanian:2020jhl}.

One manifestation of the emergence of ensemble averages in the presence of wormholes is the non-factorization of partition functions. Let us consider the simplest example, where we have two  disjoint conformal boundaries. Then the gravitational path integral receives  contributions from wormholes connecting the two boundaries.  This is puzzling from the boundary field theory point of view, because on  these two disjoint boundaries, its partition function gets factorized \cite{Maldacena:2004rf}. The resolution is to regard its dual as an ensemble of theories living on the boundaries. 
The relation between  ensemble averages and the physics of wormholes was extensively discussed a few decades ago \cite{Coleman:1988cy, Giddings:1988cx},
and recent discussions mainly focus on its implications for holographic duality
\cite{Saad:2018bqo, Maldacena:2017axo, Maldacena:2018lmt, Saad:2019pqd, Giddings:2020yes, Belin:2020hea, Cotler:2020ugk, Garcia-Garcia:2020ttf, Marolf:2021kjc, Saad:2021rcu, Verlinde:2021kgt, Verlinde:2021jwu}
and the black hole information paradox \cite{Penington:2019kki, Almheiri:2019qdq, Pollack:2020gfa, Chen:2020tes, Liu:2020jsv, Stanford:2020wkf, Goto:2020wnk}.
  
Given these developments, a natural question arises: How do we get  a bulk gravity description of a fixed boundary field theory {\it without averaging} ?
We are particularly interested in the fate of Euclidean wormholes in such a setup. 
There have been several attempts toward this goal.
For example, it was shown in \cite{Blommaert:2019wfy} that by introducing ``eigenbranes'' in the bulk of JT gravity,
one can fix several eigenvalues of the dual random matrix.
This question was studied in \cite{Saad:2021rcu} for the SYK model with one time point (i.e., zero-dimensional SYK model) with a {\it fixed} coupling constant.
There, in the calculation of a product of partition functions $Z_{L} Z_{R}$ without averaging, two types of large $N$ saddles are found.
One is the usual wormhole saddles which connect two boundaries $L$ and $R$, and give the averaged two-point function $ \la Z_{L} Z_{R} \ra$ over the random coupling. 
In addition to these,  there is another type of saddles called ``half-wormholes''.
From a bulk gravity point of view, these saddles can be interpreted as geometries starting from a conformal boundary and ending somewhere in the bulk.
Quite remarkably, the contributions of these half-wormholes are essential for $Z_{L} Z_{R}$ to get factorized. This idea is further developed in \cite{Mukhametzhanov:2021nea, Blommaert:2021gha, Garcia-Garcia:2021squ, Saad:2021uzi, Okuyama:2021eju, Goto:2021mbt, Mukhametzhanov:2021hdi, Blommaert:2021fob} for more details.

To further investigate these ideas, wormholes vs. half-wormholes and ensemble averaging vs. fixed-coupling (or theory),
in this paper, we introduce a partially disorder-averaged SYK model.
This model has a real parameter $\s$ that smoothly interpolates
between the ordinary totally disorder-averaged SYK model ($\s = \inf$) and the totally fixed-coupling model ($\s =0$).
This is archived by modifying the probability distribution of the coupling constant.
In the $\s = \inf$ limit, the probability distribution is reduced to the ordinary Gaussian form. In contrast, in the $\s =0$ limit, the probability distribution becomes a product of delta functions that enforces each component of the coupling constant to a fixed external value.
This type of interpolation between the totally disorder-averaged and fixed-coupling cases is previously studied in the context of matrix integrals in \cite{Blommaert:2021gha}.

One convenient way to study the conventional SYK model in the large $N$ limit is introducing bi-local collective fields $\{G(\t_{1}, \t_{2}), \Sigma (\t_{1}, \t_{2}) \}$
depending on two boundary times.
This formalism has a further advantage for studies of two-point (or higher-point) functions of the partition functions.
When each time corresponds to each boundary theory, the bi-local fields have a natural bulk interpretation for the corresponding gravitational configurations,
including Euclidean wormholes in a suitable setup.
Below we will argue that in the large $N$ analysis of our deformed SYK model,
we need to introduce additional local collective fields $\{G_{\sigma}(\t), \Sigma_{\sigma}(\t) \}$.
These local fields are analogs of ``half-wormholes'' because they depend only on one boundary time.
We will study in detail the behavior of these local fields $\{G_{\sigma}(\t), \Sigma_{\sigma}(\t)\}$ as we change the deformation parameter $\sigma$,
from $\sigma=\infty$ whose low energy part has a complete gravitational description in terms of JT gravity,
to the totally fixed coupling model $\sigma= 0$ whose gravitational description we would like to know better.

\bigskip

The remainder of the paper is organized as follows.
In section~\ref{sec:partial disorder-averaging}, we define a partially disorder-averaged SYK model
by modifying the probability distribution of the coupling constants.
We show that this correctly interpolates between the totally disorder-averaged and totally fixed coupling case.

In section~\ref{sec:partition function}, we study one-point function of the partition function after this partial disorder averaging.
For the large $N$ effective description, in addition to the usual bi-local collective fields, we also introduce a new additional set of local collective fields.
We explain that these local fields can be understood as the ``half'' of the bi-local collective fields.
Before studying the one-dimensional model, in section~\ref{sec:0d SYK1}, we study the zero-dimensional model (i.e., SYK model with one time point).
We show that the partially disorder-averaged partition function is proportional to the ``hyperpfaffian'' of the external coupling
together with a coefficient that vanishes in the total disorder-averaged limit while giving a finite value in the total fixed coupling limit.
Then, we go back to the one-dimensional model. After reviewing the totally disorder-averaged case
(which is well-studied in the literature) in section~\ref{sec:otal disorder-averaging}, 
we study the slightly fixed coupling case in section~\ref{sec:slightly fixed-coupling}.
We show that for the large $N$ collective field description,
the effect of slightly fixed coupling is given by the same form as the chemical potential contribution in the complex SYK model.
In section~\ref{sec:bulk interpretation}, we explain a possible bulk picture of the new local collective fields
and in section~\ref{sec:general sigma} we comments on the general $\s$ (i.e. general partial disorder-averaged) case.

In section~\ref{sec:2pt function}, we study two-point function of the partition functions after the partial disorder averaging.
In section~\ref{sec:0d SYK2}, focusing on the zero-dimensional model, we study the wormhole and half-wormhole saddles in detail. 
We also comment on the case when we add explicit interactions between the left and right SYK's in section~\ref{sec:adding a coupling}.
Our discussion on the Brownian SYK model is given in section~\ref{sec:brownian}.

We study the transition of the spectral density as well as the spectral form factor from the totally fixed coupling limit to the totally disorder-averaged limit in section \ref{sec:spectrum}.
Our conclusions and further discussions are presented in section~\ref{sec:conclusions}.
Supplemental computations and discussion are gathered in appendices.

\section{Partial Disorder A\hspace{-0.6mm}verage}
\label{sec:partial disorder-averaging}
In this section, we define our partially disorder-averaged SYK model,
which smoothly interpolates between the ordinary totally disorder-averaged SYK model and the totally fixed coupling model by introducing a real parameter $\sigma$.  
The original SYK model \cite{Sachdev:1992fk, Kitaev:2015} is defined by the Hamiltonian 
	\begin{align}
		H \, = \, i^{\frac{q}{2}} \sum_{i_1<i_2<\cdots<i_q}^N J_{i_1 \cdots i_q} \, \c_{i_1} \cdots \c_{i_q} \, ,
 	\end{align}
where the one-dimensional Majorana fermions satisfy 
	\begin{align}
		\{ \c_i, \c_j \} \, = \, \d_{ij} \, .
 	\label{ACR}
	\end{align}
The coupling $J_{i_{1} \cdots i_{q}}$ is  random, 	 and  drawn from Gaussian ensemble with the vanishing mean  value $\la J_{i_{1} \cdots i_{q}} \ra=0 $.
Its Lagrangian is given by
	\begin{align}
		L \, = \, \frac{1}{2} \sum_{i=1}^N \c_i \pa_\t \c_i \, - \, i^{\frac{q}{2}} \sum_{i_1<\cdots<i_q}^N J_{i_1 \cdots i_q} \, \c_{i_1} \cdots \c_{i_q} \, .
 	\label{Lagrangian}
 	\end{align}

Now we introduce a partial disorder averaging by deforming the probability distribution of the random coupling as
	\begin{align}
		P(J_{i_1 \cdots i_q}) \, = \, \exp\left[ -\frac{N^{q-1}}{2(q-1)!} \sum_{i_1<\cdots<i_q}^N
		\left( \frac{J_{i_1 \cdots i_q}^2}{J^2} \, + \, \frac{(J_{i_1 \cdots i_q} - J_{i_1 \cdots i_q}^{(0)})^2}{\s^2} \right)\right] \, ,
 	\label{P(J)}
 	\end{align}
with 
	\begin{align}
		\big\la \mathcal{O} \big\ra_J \, \equiv \, \mathcal{N}_\s^{-1} \int \prod_{i_1<\cdots<i_q}^N dJ_{i_1 \cdots i_q} \, P(J_{i_1 \cdots i_q}) \, \mathcal{O} \, .
 	\end{align}
Here $J_{i_1 \cdots i_q}^{(0)}$ is the external fixed coupling, and the parameter $\s$ controls how much we take disorder averaging.
This type of partial disorder averaging was studied in \cite{Blommaert:2021gha} for a matrix model.
As in \cite{Blommaert:2021gha}, $\s = \inf$ corresponds to the total disorder-averaging coincides with the ordinary SYK model,
while $\s \to 0$ corresponds to totally fixing the coupling constant $J_{i_1 \cdots i_q}= J_{i_1 \cdots i_q}^{(0)}$.
In the following, it is convenient to introduce a new set of effective couplings by
	\begin{align}
		\frac{1}{\wtd{J}^2} \, \equiv \, \frac{1}{J^2} \, + \, \frac{1}{\s^2} \, , \qquad J_\s \, \equiv \, \frac{\wtd{J}^2 J_0}{\s^2} \, ,
		\label{Js}
  	\end{align}
where $J_0$ is defined in (\ref{J_0}) and we assume $\wtd{J}\ge0$.
In the extreme limits of the parameter $\s$, these two couplings behave as
	\begin{align}
		\s \, \to \, \inf \, :& \qquad \wtd{J} \, \to \, J \, , \qquad J_\s \, \to \, 0 \, , \\
		\s \, \to \, 0 \, :& \qquad \wtd{J} \, \to \, \s \, , \qquad J_\s \, \to \, J_0 \, .
  	\label{Jtilde}
  	\end{align}
The normalization factor $\mathcal{N}_\s$ is defined by
	\begin{align}
		\mathcal{N}_\s \, \equiv \, \int \prod_{i_1<\cdots<i_q}^N dJ_{i_1 \cdots i_q} \, P(J_{i_1 \cdots i_q}) \, ,
 	\end{align}
and this gives us
	\begin{align}
		\mathcal{N}_\s^{-1} \, = \, \left( \frac{N^{q-1}}{2(q-1)! \pi \wtd{J}^2} \right)^{\frac{1}{2} \binom{N}{q}} \,
		\exp \left( \frac{N^{q-1}}{2(q-1)!(J^2+\s^2)} \sum_{i_1<\cdots<i_q}^N \Big( J_{i_1 \cdots i_q}^{(0)} \Big)^2 \right) \, .
 	\end{align}
For the usual total-disorder-averaged SYK model, this factor does not play any crucial role,
but for our partially disorder-averaged SYK model, we have to keep at least nontrivial $\s$-dependence coming from this normalization factor.
For example, in the totally-fixed-coupling limit ($\s \to 0$), combined with this normalization factor, the probability distribution gives delta functions:
	\begin{align}
		\lim_{\s \to 0} \ \mathcal{N}_\s^{-1} \, P(J_{i_1 \cdots i_q}) \, = \, \prod_{i_1<\cdots<i_q}^N \d\left(J_{i_1 \cdots i_q} - J_{i_1 \cdots i_q}^{(0)}\right) \, .
 	\end{align}
This type of delta function projection of the coupling constant was also studied in \cite{Mukhametzhanov:2021hdi}.

This definition of partial disorder averaging does not give a vanishing mean value of the coupling unlike the ordinary SYK model, but we find
	\begin{align}
		\big\la J_{i_1 \cdots i_q} \big\ra_J \, = \, \frac{\wtd{J}^2}{\s^2} \, J_{i_1 \cdots i_q}^{(0)} \, , \qquad
		\big\la J_{i_1 \cdots i_q}^2 \big\ra_J \, = \, \frac{(q-1)!\wtd{J}^2}{N^{q-1}} \left( 1 + \frac{\wtd{J}^2 N^{q-1}}{(q-1)!\s^4} \, \big(J_{i_1 \cdots i_q}^{(0)}\big)^2 \right) \, .
 	\label{<J_i>}
 	\end{align}
In the total disorder-averaging limit, we recover the ordinary SYK results:
$\big\la J_{i_1 \cdots i_q} \big\ra_J^{\s=\inf} = 0$ and $\big\la J_{i_1 \cdots i_q}^2 \big\ra_J^{\s=\inf} = (q-1)!J^2/N^{q-1}$,
while in the totally fixed-coupling limit, we find
$\big\la J_{i_1 \cdots i_q} \big\ra_J^{\s=0} = J^{(0)}_{i_1 \cdots i_q}$ and $\big\la J_{i_1 \cdots i_q}^2 \big\ra_J^{\s=0} = \big( J^{(0)}_{i_1 \cdots i_q} \big)^2$.

\section{Partially Disorder-A\hspace{-0.6mm}veraged Partition Function}
\label{sec:partition function}

Having specified the model of our interest, in this section we study the partially disorder-averaged partition function of this model in the large $N$ limit, 
by introducing collective fields. We see that to study the partially disordered model with finite $\s$,  in addition to ordinary bi-local fields $\{ G(\tau_{1}, \tau_{2}), \Sigma(\tau_{1}, \tau_{2})\}$ in the totally averaged theory at $\sigma= \infty$,   it is convenient to introduce  
an additional set of collective fields denoted by $\{G_{\sigma}(\tau), \Sigma_{\sigma}(\tau)\}$. 
We will explain a possible bulk picture of these local collective fields in section~\ref{sec:bulk interpretation}
and for the two-point function we study in section~\ref{sec:2pt function}, they eventually correspond to the degrees of freedom of ``half-wormhole''.

The original partition function with fixed coupling $J_{i_1 \cdots i_q}$ is given by 
	\begin{align}
		Z \, &= \, \int D\c_i \, e^{-S} \nn\\
		\, &= \, \int D\c_i \, \exp\left[ - \frac{1}{2} \int d\t \sum_{i=1}^N \c_i \pa_\t \c_i \, + \, i^{\frac{q}{2}} \int d\t \sum_{i_1<\cdots<i_q}^N J_{i_1 \cdots i_q} \, \c_{i_1} \cdots \c_{i_q} \right] \, .
 	\end{align}
Taking the partial disorder averaging of this partition function with respect to the probability distribution \eqref{P(J)}, we obtain
	\begin{align}
		\big\la Z \big\ra_J \, = \, \int D\c_i \, e^{-S_{\rm eff}[\c]} \, ,
		\label{zJ}
 	\end{align}
where
	\begin{align}
		S_{\rm eff}[\c] \, &= \, \frac{1}{2} \int d\t \sum_{i=1}^N \c_i \pa_\t \c_i
		\, - \, \frac{\wtd{J}^2}{2qN^{q-1}} \int d\t_1 d\t_2 \left( \sum_{i=1}^N \c_i(\t_1) \c_i(\t_2) \right)^q \nn\\
		&\quad - \, \frac{i^{\frac{q}{2}}\wtd{J}^2}{\s^2} \int d\t \sum_{i_1<\cdots<i_q}^N J_{i_1 \cdots i_q}^{(0)} \, \c_{i_1} \cdots \c_{i_q} \, .
 	\label{S_eff[chi]}
 	\end{align}
The normalization factor $\mathcal{N}_\s$ is canceled out by the Gaussian integral of the coupling.	
We emphasize that even though we integrated $J_{i_1 \cdots i_q}$ to obtain $\la Z \ra_J$, 
this still can be the totally fixed-coupling partition function in the $\s \to 0$ limit.
Indeed, we can see that in the totally fixed-coupling limit ($\s \to 0$), the second term vanishes,
while the last term survives because $\wtd{J}^2/\s^2 \to 1$.
The result is nothing but the totally fixed-coupling action (\ref{Lagrangian}).
On the other hand, in the totally disorder-averaged limit ($\s \to \inf$) the last term vanishes
and the remaining first two terms are the usual disorder-averaged SYK effective action.

We also note that besides the last term, the original SYK effective action has the global $O(N)$ gauge symmetry: $\c_i \to \sum_j O_{ij} \c_j$.
However, this global $O(N)$ symmetry is now broken by the last term by fixing the coupling constant $J_{i_1 \cdots i_q}^{(0)}$.
In addition, this term contains a factor $i^{q/2}$, so in order to keep the action real, it looks like we need a complex $J_{i_1 \cdots i_q}^{(0)}$,
such that the entire last term is still real.
This type of complex coupling is similar to the one discussed in \cite{Garcia-Garcia:2020ttf, Garcia-Garcia:2021squ}, but the precise relation is unclear.\\

\bigskip 
{\bf Fixed-couplings as an external source}\\
Here we will explain how to introduce collective fields in our partially disorder-averaged SYK model.
For the first line in (\ref{S_eff[chi]}), we introduce a set of bi-local fields $(G_{LR},\Sigma_{LR})$ as usual.
To introduce a new set of collective fields for the second line in (\ref{S_eff[chi]}), we will give the following interpretation of the external fixed couplings.
	
Let us now assume a specific form of the external coupling $J_{i_1 \cdots i_q}^{(0)}$.
We choose it to be proportional to the totally anti-symmetric tensor $\varepsilon_{{i_{1},\cdots i_{q}}}$ for $i_k=1,\cdots, N$, with $\varepsilon_{1,2,\cdots, q}=1 $, $\varepsilon_{2,3,\cdots, q+1}=1 $, e.t.c.
From now on, we interpret the second term in (\ref{S_eff[chi]}) as the couplings between the fermions $\chi_i$ in the SYK model and the external fermionic  sources $\theta_i$ by decomposing the anti-symmetric tensor $\varepsilon_{{i_{1},\cdots i_{q}}}$ into the products of $\theta_i$,
	\begin{align}
		\qquad J_{i_1 \cdots i_q}^{(0)} \, = \, \frac{(q-1)! J_0}{i^{\frac{q}{2}} N^{q-1}} \, \th_{i_1} \cdots \th_{i_q} \, , \qquad \quad {\rm with} \qquad \{\th_i, \th_j\} \, = \, \d_{ij} \, ,
 	\label{J_0}
	\end{align}
where the coefficient was chosen for later convenience.
\footnote{Here we introduced a non-vanishing square of the variables $\th_i$ in order to keep the second term of $\la J_{i_1 \cdots i_q}^2 \ra_J$ in (\ref{<J_i>}) non-zero
(or to keep non-vanishing $\la J_{i_1 \cdots i_q}^2 \ra_J^{\s=0}$).
If we regard $\th_i$ as external Grassmann variables, it is natural to have $\{\th_i, \th_j\}=0$,
but we can think that the external coupling originally had a slight time-dependence, and we smeared over that time-dependence, so we have $\th_i^2=1/2$.
A similar argument was also given in footnote~9 of \cite{Saad:2018bqo}.
}

This variable $\th_i$ can be interpreted as non-dynamical fermions living in another universe. We discuss this interpretation in Appendix~\ref{app:external coupling}.
After using this decomposition, we have the effective action
	\begin{align}
		S_{\rm eff}[\c] \, &= \, \frac{1}{2} \int d\t \sum_{i=1}^N \c_i \pa_\t \c_i
		\, - \, \frac{\wtd{J}^2}{2qN^{q-1}} \int d\t_1 d\t_2 \left( \sum_{i=1}^N \c_i(\t_1) \c_i(\t_2) \right)^q \nn\\
		&\quad - \, \frac{J_\s}{qN^{q-1}} \int d\t \left( \sum_{i=1}^N \th_i \c_i(\t) \right)^q \, ,
	\label{seff}
 	\end{align}
with $J_{\sigma}$ defined in \eqref{Js}. Now the term depending on the fixed couplings is rewritten as a local term similar to the usual bi-local term in the first line, which enables us to introduce the collective fields to this term.  Now we are considering the SYK model coupled to the external fields $\theta_i$. We will analyze this model instead of the original partially disorder-averaged SYK model. After all computations, we replace all the dependence of $\theta_i$ by the original fixed-couplings $J_{i_1 \cdots i_q}^{(0)}$, then we expect to recover the result in the original partially disorder-averaged SYK model.  

Now it is useful to employ the following Hubbard–Stratonovich trick:
	\begin{align}
		1 \, &= \, \int DG \int DG_\s \, \d\left( G - \frac{1}{N} \sum_{i=1}^N \c_i \c_i \right)
		\d\left( G_\s - \frac{1}{N} \sum_{i=1}^N \th_i \c_i \right) \\
		&= \, \int DG D\S DG_\s D\S_\s \exp\left[ \frac{1}{2} \Tr \, \S \cdot \left( N G - \sum_{i=1}^N \c_i \c_i \right)
		- \Tr \, \S_\s \cdot \left( N G_\s - \sum_{i=1}^N \th_i \c_i \right) \right] \, , \nn
 	\label{HS-trick}
	\end{align}
We insert the above identity to the partially disorder-averaged partition function \eqref{zJ} with $S_{\rm eff}[\c]$ given by \eqref{seff}.
Rewriting all terms in $S_{\rm eff}[\c]$ \eqref{seff} by the bi-local $G$ and the local $G_\s$ fields,
we can perform the fermion integrals (see Appendix~\ref{app:integration} for detail) and obtain
	\begin{align}
		\big\la Z \big\ra_J \, = \, \mathcal{N}_\s^{-1} \int DG D\S DG_\s D\S_\s \, e^{-S_{\rm eff}[G, \S, G_\s, \S_\s]} \, 
 	\end{align}
with
	\begin{align}
		&S_{\rm eff}[G, \S, G_\s, \S_\s] \, = \, - \frac{N}{2}\Tr\log(\S ) - \frac{N}{2} \int d\t \big[ \pa_\t G(\t, \t')\big]_{\t'=\t} \nn\\
		&\qquad \qquad \qquad  +\frac{N}{2} \int d\t_1 d\t_2 \left( \S(\t_1, \t_2) G(\t_1, \t_2) - \frac{\wtd{J}^2}{q} G(\t_1, \t_2)^q \right)  \label{S_eff} \\
		&\qquad + N \int d\t \left( \S_\s(\t) G_\s(\t) - \frac{J_\s}{q} G_\s(\t)^q \right)
		\, - \, \frac{N}{4} \int d\t_1 d\t_2 \, \S_\s(\t_1) G(\t_1, \t_2) \S_\s(\t_2) \, . \nn
	\end{align}

Before studying this effective action in detail, let us first consider the meaning of the local field $G_\s$ we introduced here in addition to the usual SYK bi-local fields.
If we consider the combination of $G_\s^q(\t_1) G_\s^q(\t_2)$, this is written as
	\begin{align}
		\Big( G_\s(\t_1) G_\s(\t_2) \Big)^q \, = \, \frac{i^q}{\big((q-1)!\big)^2 J_0^2 N^2} \,
		&\sum_{i_1, \cdots ,i_q}^N J_{i_1 \cdots i_q}^{(0)} \, \c_{i_1}(\t_1) \cdots \c_{i_q}(\t_1) \nn\\
		&\times \sum_{j_1, \cdots ,j_q}^N J_{j_1 \cdots j_q}^{(0)} \, \c_{j_1}(\t_2) \cdots \c_{j_q}(\t_2) \, ,
 	\end{align}
where we used the decomposition of $J_{i_1 \cdots i_q}^{(0)}$ (\ref{J_0}).
If we take the total disorder averaging of $J_{i_1 \cdots i_q}^{(0)}$ for this quantity
with using the same form of probability distribution as (\ref{P(J)}) for $J_{i_1 \cdots i_q}^{(0)}$, we find 
	\begin{align}
		\Big\la G_\s^q(\t_1) G_\s^q(\t_2) \Big\ra_{J_0} \, = \, \frac{i^q \wtd{J}^2}{(q-1)! J_0^2 N} \, G(\t_1, \t_2)^q \, .
 	\end{align}
This relation implies that we can think of the local field $G_\s(\t)$ as the ``half'' of the bi-local field $G(\t_1, \t_2)$. Furthermore, one can regard the anti-symmetric variable  $\theta_{i}$ as fermionic degrees of freedom living in the surface at which the  bulk spacetime ends. Since the new variable  $G_{\s} (\tau)$ is  the propagator between the  original SYK fermions $\c_{i}$ and the new ones $\theta_{i}$, this also suggests that $G_{\s} (\tau)$ corresponds to the correlation between the boundary and the  interior of the bulk. These interpretations will be useful in the following discussion.

Since all terms in the effective action (\ref{S_eff}) are proportional to $N$, we can use large $N$ saddle-point evaluation. Variation of each field leads to the following saddle-point equations
	\begin{align}
		\d G \, :& \qquad \S(\t_1, \t_2) \, = \, \wtd{J}^2 \, G(\t_1, \t_2)^{q-1}
		\, + \, \pa_{\t_1} \d(\t_1 - \t_2) \, + \, \frac{1}{2} \, \S_\s(\t_1) \S_\s(\t_2)\, , \label{delta G} \\
		\d \S \, :& \qquad \S^{-1}(\t_2, \t_1) \, = \, G(\t_1, \t_2) \, , \label{delta Sigma} \\
		\d G_\s \, :& \qquad \S_\s(\t) \, = \, J_\s \, G_\s(\t)^{q-1} \, , \label{delta G_sigma} \\
		\d \S_\s \, :& \qquad G_\s(\t) \, = \, \frac{1}{2} \int d\t' \, G(\t, \t') \S_\s(\t') \, , \label{delta Sigma_sigma}
  	\end{align}
where the inverse is defined by 
	\begin{align}
		\int d\t_3 \, \S(\t_1, \t_3) \S^{-1}(\t_3, \t_2) \, = \, \d(\t_1 - \t_2) \, .
  	\label{inverse}
  	\end{align}
From (\ref{delta G}), in order to keep the time translation symmetry, the only possible solution for the local field $\S_\s$ is a time-independent constant solution.
Such a constant term in (\ref{delta G}) has the same structure as the chemical potential contribution in the complex SYK model
as we will discuss more in detail in section \ref{sec:slightly fixed-coupling}.

\subsection{Zero-dimensional SYK}
\label{sec:0d SYK1}
Before studying this model in detail, let us first consider the zero-dimensional case discussed in \cite{Saad:2021rcu, Mukhametzhanov:2021nea}.
This model is not dynamical, but it might capture some important topological contributions.
The treatment of this case is slightly different from the one-dimensional case we discussed above.
This is because in the effective action (\ref{S_eff[chi]}), the second term identically vanishes in this case,
and of course, the kinetic term also does not exist:
	\begin{align}
		\big\la Z \big\ra_J \, = \, \int d\c_i \, \exp\left[ \frac{J_\s}{q N^{q-1}} \bigg( \sum_{i=1}^N \th_i \c_i \bigg)^q \right] \, .
 	\end{align}
Therefore, in this case, we cannot introduce the bi-local fields, which are identically zero, so at most, we can introduce the local fields only:
	\begin{align}
		1 \, = \, \int_{-\inf}^{\inf} dG_\s \int_{-i\inf}^{i\inf} \frac{d\S_\s}{2\pi i/N} \, \exp\bigg[ \S_\s \Big( N G_\s - \sum_{i=1}^N \th_i \c_i \Big) \bigg] \, .
 	\label{HS-trick-0d}
 	\end{align}
Hence, in this case, after integrating out the fermions (see Appendix~\ref{app:integration} for detail) we have
	\begin{align}
		\big\la Z \big\ra_J \, = \, \frac{N}{2\pi i} \, \bigg( \prod_{i=1}^N \th_i \bigg) \int dG_\s d\S_\s \, e^{-S_{\rm eff}[G_\s, \S_\s]} \, ,
 	\end{align}
with 
	\begin{align}
		S_{\rm eff}[G_\s, \S_\s] \, = \, - \, N \log(-\S_\s) \, - \, N \left( \S_\s G_\s + \frac{J_\s}{q} G_\s^q \right) \, ,
 	\end{align}
where the $\log(-\S_\s)$ term comes from expanding the linear in the fermion term in the Hubbard–Stratonovich trick (\ref{HS-trick-0d}).

Now we can simply integrate out both fields as
	\begin{align}
		\big\la Z \big\ra_J \, &= \, \frac{N}{2\pi i} \, \bigg( \prod_{i=1}^N \th_i \bigg) \int dG_\s d\S_\s \, \left( - N^{-1} \pa_{G_\s} \right)^N
		e^{N \left( \S_\s G_\s + \frac{J_\s}{q} G_\s^q \right)} \nn\\
		&= \, \bigg( \prod_{i=1}^N \th_i \bigg) \int dG_\s \, e^{ \frac{N J_\s}{q} G_\s^q} \left( - N^{-1} \pa_{G_\s} \right)^N \, \d(G_\s) \nn\\
		&= \, \bigg( \prod_{i=1}^N \th_i \bigg) \left( \frac{J_\s N}{q} \right)^{\frac{N}{q}} \frac{N!}{N^N (N/q)!} \, ,
 	\label{<Z>_J-exact0}
 	\end{align}
where for the last line, we assume $N$ is a multiple of $q$.
For the total-disorder-averaging limit ($\s \to \inf$), we have $J_\s \to 0$ (\ref{Jtilde}). Therefore, in this limit we have $\la Z \ra_J = 0$.
On the other hand, for finite $\s>0$, we have a nontrivial partition function, which is proportional to the ``hyperpfaffian'' of the external coupling \cite{Mukhametzhanov:2021nea}:
	\begin{align}
		{\rm PF}\Big( J_{i_1 \cdots i_q}^{(0)} \Big) \, &\equiv \, \sideset{}{'}\sum_{A_1 < \cdots < A_p} \sgn(A) \, J_{A_1}^{(0)} \cdots J_{A_p}^{(0)} \nn\\
		&= \, \frac{N!}{p! (q!)^p} \, J_{1 \, \cdots \, q}^{(0)} \cdots J_{N-q \, \cdots \, N}^{(0)} \nn\\
		&= \, \frac{N!}{p! N^N} \left( \frac{N J_0}{q} \right)^{\frac{N}{q}} \bigg( \prod_{i=1}^N \th_i \bigg) \, ,
 	\end{align}
where $p=N/q$ and $A$'s denote ordered $q$-subsets $A={i_1< \cdots < i_q}$.
For the third line, we used (\ref{J_0}) and assumed $N \in 8\mathbb{Z}$ to eliminate the factor coming from the $i^{q/2}$.
Therefore, we find
	\begin{align}
		\big\la Z \big\ra_J \, = \, \bigg( \, \frac{\wtd{J}}{\s} \, \bigg)^{\frac{2N}{q}} {\rm PF}\Big( J_{i_1 \cdots i_q}^{(0)} \Big) \, .
 	\end{align}

Going back to the expression of (\ref{<Z>_J-exact0}), for the numerical coefficient, besides the $J_\s$ dependence, we find in the large $N$ approximation that
	\begin{align}
		\big\la Z \big\ra_J \, \approx \, \bigg( \prod_{i=1}^N \th_i \bigg) \big( J_\s \big)^{\frac{N}{q}} \sqrt{q} \, e^{-(1-\frac{1}{q}) N} \, .
 	\label{<Z>_J-exact}
 	\end{align}

We can also evaluate this partition function in the large $N$ saddle-point approximation.
From the effective action, we have the large $N$ saddle-point equations
	\begin{align}
	    \frac{1}{\S_\s} \, = \, -  \, G_\s \, , \qquad \S_\s \, = \, - \, J_\s \, G_\s^{q-1} \, ,
 	\end{align}
and there are $q$ solutions:
	\begin{align}
	    G_\s^{(s)} \, = \, e^{\frac{2m\pi i}{q}} \, J_\s^{- \frac{1}{q}} \, , \qquad \S_\s^{(s)} \, = \, - \, e^{-\frac{2m\pi i}{q}} \, J_\s^{\ \frac{1}{q}} \, ,
 	\end{align}
where $m=1,2,\cdots, q$.
Evaluating the on-shell action for these large $N$ saddle-point solutions, we find 
	\begin{align}
	    S_{\onshell} \, = \, - \, \frac{N}{q} \log J_\s \, + \, N \left( 1 - \frac{1}{q} \right) \, + \, \frac{2m \pi i N}{q} \, ,
 	\end{align}
and 
	\begin{align}
		\big\la Z \big\ra_J \, = \, \bigg( \prod_{i=1}^N \th_i \bigg) \big( J_\s \big)^{\frac{N}{q}} \, \sum_{m=1}^q e^{-(1-\frac{1}{q})N-\frac{2m \pi i N}{q}} \, .
 	\end{align}
We obtained the same $J_\s$- and $N$-dependence as the above exact computation.
The remaining sum can also be evaluated as
	\begin{align}
		\sum_{m=1}^q e^{-2mp \pi i} \, = \, \frac{1-e^{2pq\pi i}}{e^{2pq\pi i}(1-e^{2p \pi i})} \, = \, q \, ,
 	\label{m-sum}
 	\end{align}
where we put $p=N/q$ and for the first equality we assumed $p\ne \mathbb{Z}$, but for the second equality we took $p\to \mathbb{Z}$ and evaluated by using l'H\^{o}pital's rule.
The $1/\sqrt{q}$ difference from the exact result (\ref{<Z>_J-exact}) is accounted by the one-loop contribution \cite{Saad:2021rcu}.
To evaluate this contribution, we expand the fields around the saddle-point solutions:
	\begin{align}
	    G_\s \, = \, G_\s^{(s)} \, + \, \frac{\d G_\s}{\sqrt{N}} \, , \qquad \S_\s \, = \, \S_\s^{(s)} \, + \, \frac{\d \S_\s}{\sqrt{N}} \, .
 	\end{align}
Then, the quadratic action is written as
	\begin{align}
	    S_{(2)} \, &= \, \frac{1}{2} \Big[ J_\s^{-\frac{2}{q}} e^{\frac{4m\pi i}{q}} \, \d\S_\s^2 \, - \, 2 \d\S_\s \d G_\s
	    \, - \, (q-1) J_\s^{\ \frac{2}{q}} e^{-\frac{4m\pi i}{q}} \, \d G_\s^2 \Big] \nn\\
	    &= \, \frac{1}{2} \Big[ \d\tilde{\S}_\s^2 \, + \, 2i \d\tilde{\S}_\s \d \tilde{G}_\s \, + \, (q-1) \d \tilde{G}_\s^2 \Big] \, ,
 	\end{align}
where in the second line we change the variables by $\tilde{G}_\s = i J_\s^{1/q} e^{-\frac{2m\pi i}{q}} \d G_\s$ and 
$\tilde{\S}_\s = J_\s^{-1/q} e^{\frac{2m\pi i}{q}} \d\S_\s$ to make the Gaussian integral convergent.
Therefore, the one-loop contribution is given by $\big\la Z \big\ra_J^{\oneloop} = 1/\sqrt{q}$. Finally, including this one-loop contribution, we find 
	\begin{align}
		\big\la Z \big\ra_J \, = \, \bigg( \prod_{i=1}^N \th_i \bigg) \big( J_\s \big)^{\frac{N}{q}} \sqrt{q} \, e^{-(1-\frac{1}{q}) N} \, .
 	\end{align}
This precisely agrees with (\ref{<Z>_J-exact}).

\subsection{Total disorder-averaging}
\label{sec:otal disorder-averaging}
Now let us come back to the one-dimensional model \eqref{S_eff}.
In the totally disorder-averaging limit ($\s \to \inf$), we have $J_\s \to 0$.
This means that the variation of $G_\s$ (\ref{delta G_sigma}) sets $\S_\s = 0$. 
Therefore, the remaining dynamics is described by the bi-local fields $\{ G, \S\}$, which is nothing but the effective system of the ordinary SYK model.

Combining (\ref{delta G}) and (\ref{delta Sigma}), the Schwinger-Dyson equation for $G(\t_1, \t_2)$ is now obtained as
	\begin{align}
	    - \d(\t_1 - \t_2) \, &= \, \int d\t_3 \, G(\t_1, \t_3) \S(\t_3, \t_2) \nn\\
	    &= \, \pa_{\t_1} G(\t_1, \t_2) \, + \, J^2 \int d\t_3 \, G(\t_1, \t_3) G(\t_3, \t_2)^{q-1} \, .
  	\end{align}
The free solution, which is obtained by setting $J=0$, is found as
	\begin{align}
	    G_{\rm free}(\t_1, \t_2) \, = \, \frac{\sgn(\t_{12})}{2} \, ,
  	\end{align}
where $\t_{12} \equiv \t_1 - \t_2$.
On the other hand, the IR solution, which is obtained by eliminating the kinetic term, is given by
	\begin{align}
	    G^\IR_{\b = \inf}(\t_1, \t_2) \, = \, \frac{b^\D}{J^{2\D}} \, \frac{\sgn(\t_{12})}{|\t_{12}|^{2\D}} \, ,
  	\label{GIR_b=inf}
  	\end{align}
for zero temperature case where $\D = 1/q$ and 
	\begin{align}
	    b^\D \, = \, \left( \frac{1-2\D}{2}\right) \, \tan(\pi \D) \, .
  	\end{align}
This solution can be determined by using the scaling ansatz and the Fourier transform
	\begin{align}
	    \frac{\sgn(\t)}{|\t|^\a} \, = \, c(\a) \int \frac{d\o}{2\pi} \, e^{- i \o \t} \, |\o|^{\a-1} \sgn(\o) \, , \qquad
	    c(\a) \, = \, i 2^{1-\a} \sqrt{\pi} \, \frac{\G(1-\frac{\a}{2})}{\G(\frac{1}{2}+\frac{\a}{2})} \, .
  	\end{align}

Now we realize that the IR effective action
	\begin{align}
		S_{\rm IR}[G] \, = \, \frac{N}{2}\Tr\log(G ) - \frac{NJ^2}{2q} \int d\t_1 d\t_2 \, G(\t_1, \t_2)^q \, ,
	\label{S_IR}
	\end{align}
is invariant under the conformal transformation $\t \to f(\t)$ together with the bi-local field transformation
	\begin{align}
		G(\t_1, \t_2) \, \to \, G_f(\t_1, \t_2) \, \equiv \, |f'(\t_1) f'(\t_2)|^{\D} \, G(f(\t_1), f(\t_2)) \, .
	\label{G_f}
	\end{align}
For the IR solution (\ref{GIR_b=inf}), a general transformation gives
	\begin{align}
	    G^\IR_f(\t_1, \t_2) \, = \, \frac{b^\D}{J^{2\D}} \, \frac{|f'(\t_1) f'(\t_2)|^\D}{|f(\t_1) - f(\t_2)|^{2\D}} \, \sgn(\t_{12}) \, .
	\label{G^IR_f} 
 	\end{align}
In particular, the finite temperature solution can be obtained by this transformation with $f(\t) = (\pi/\b) \tan(\pi \t /\b)$ as
	\begin{align}
	    G^\IR_\b(\t_1, \t_2) \, = \, \frac{b^\D}{J^{2\D}} \, \left| \frac{\pi}{\b \sin\frac{\pi \t_{12}}{\b}} \right|^{2\D} \, \sgn(\t_{12}) \, .
  	\end{align}

\subsection{Slightly fixed-coupling}
\label{sec:slightly fixed-coupling}

In this subsection, we study the effect of slightly fixing the coupling constant.
For a slightly fixed coupling (which means large $\s$), we have
	\begin{align}
	    \wtd{J}^2 \, &= \, \frac{J^2 \s^2}{J^2 + \s^2} \, = \, J^2 \left[ 1 \, - \, \frac{J^2}{\s^2} \, + \, \mathcal{O}(\s^{-4}) \right] \, , \\
	    J_\s \, &= \, \frac{J^2 J_0}{J^2 + \s^2} \, = \, \frac{J^2J_0}{\s^2} \, + \, \mathcal{O}(\s^{-4}) \, .
  	\end{align}

Let us first consider the free limit $T\gg J$, where $T$ is temperature. We would like to keep small $J_{\s}$ corrections while totally neglect $\wtd{J}$ contribution.
This means that we have
    \begin{align}
        J_0 \, \sim \, \s \, \gg \, T \, \gg \, J \, .
    \end{align}
Therefore, we can only keep $J_{\s}$ corrections and ignore the terms proportional to $\wtd{J}^2$.
In this limit, we start from the zero-th order (in $J_\s$), where the free bi-local sector is given by  $\S(\t_1, \t_2) \, = \, \pa_{\t_1} \d(\t_{12})$ and $\S_\s = 0$.
To get the first order, inverting the equation (\ref{delta Sigma_sigma}), we find
	\begin{align}
		\S_\s(\t) \, &= \, - 2 \int d\t' \, \S(\t, \t') G_\s(\t') \nn\\
		&= \, - 2 \, \pa_\t G_\s(\t) \, .
  	\label{Sigma_sigma-eqsfc}
  	\end{align}
Therefore, equating with (\ref{delta G_sigma}), the differential equation for $G_\s(\t)$ is found as
	\begin{align}
	    \pa_\t G_\s(\t) \, = \, - \, \frac{J_\s}{2} \, G_\s(\t)^{q-1} \, ,
  	\end{align}
and the solution of this equation is 
	\begin{align}
	    G_\s(\t) \, &= \, \left( \frac{2}{(q-2)(J_\s \t+c)} \right)^{\frac{1}{q-2}} \nn\\
	    &= \, \left( \frac{2}{c(q-2)} \right)^{\frac{1}{q-2}} \left[ 1 \, - \, \frac{J_\s \t}{c(q-2)} \, + \, \mathcal{O}(J_\s^2) \right] \, ,
  	\end{align}
where $c$ is an integration constant.
From (\ref{Sigma_sigma-eqsfc}), we also find
	\begin{align}
	    \S_\s(\t) \, &= \, - \, J_\s \left( \frac{2}{(q-2)(J_\s \t+c)} \right)^{-\frac{q-1}{q-2}} \nn\\
	    &= \, - \, \left( \frac{2}{c(q-2)} \right)^{-\frac{q-1}{q-2}} \Big[ J_\s \, + \, \mathcal{O}(J_\s^2) \Big] \, .
  	\end{align}
Focusing on the leading contribution (in $J_\s$) for this $\S_\s$, the last term in (\ref{delta G}) gives a constant contribution for $\S(\t_1, \t_2)$.
Fourier transforming the bi-local time to frequency space, such a constant contribution gives a contribution $\S(\o) = \m \d(\o) \, + \, \cdots$,
where the ellipsis denotes contributions from other terms.
Such a contribution localized at $\o=0$ is the same structure as a chemical potential of the complex SYK model 
\footnote{Here we emphasize that we are just talking about the structural similarity with the complex SYK model,
but we are {\it not} claiming the model becomes the actual complex model.}
\cite{Sachdev:2015efa, Davison:2016ngz, Gu:2019jub}.
Hence, the saddle-point equation for $G(\t_1, \t_2)$ becomes
	\begin{align}
	    - \, \d(\t_1 - \t_2) \, = \, \big( \pa_{\t_1} - \m \big) G(\t_1, \t_2) \, .
  	\end{align}
This is nothing but Green's equation for a free fermion with a chemical potential $\m$, and the solution is given by
	\begin{align}
	    G_{\rm free}(\t_1, \t_2) \, = \, - \, \frac{\sgn(\t_{12}) \, e^{\m |\t_{12}|}}{1+e^{\, \sgn(\t_{12}) \b \m}} \, ,
  	\label{G_free}
  	\end{align}
where $\b$ is the inverse temperature.

We suppose the above chemical potential contribution also influences on the IR region as in the complex SYK model.
For the IR region, we stay at the strong coupling limit $J \gg T$.
On top of that, we would like to keep the leading term of $J_\s$ as a small parameter and neglect the sub-leading contribution of $\wtd{J}$.
All this means that in this subsection we study the regime of 
	\begin{align}
	    \s \, \gg \, J_0 \, \sim \, J \, \gg \, T \, . 
  	\end{align}
Including the potential contribution, the saddle-point equation for $G(\t_1, \t_2)$ can be written as
	\begin{align}
	    - \d(\t_1 - \t_2) \, = \, \big( \pa_{\t_1} - \m \big) G(\t_1, \t_2) \, + \, J^2 \int d\t_3 \, G(\t_1, \t_3) G(\t_3, \t_2)^{q-1} \, .
  	\end{align}
This equation is the saddle-point equation of the complex SYK model with a chemical potential $\m$.
We will show that the same structure also appears in another formalism of the slightly fixed coupling SYK model in Appendix~\ref{app:schwarzian}.
It is very interesting that even though we started with Majorana fermions, after slightly fixing the random couplings, we obtain the {\it induced} chemical potential.
This might be related to the branes discussed in \cite{Saad:2019lba, Blommaert:2019wfy, Okuyama:2021eju} in order to fix the boundary theory.

In the IR limit, the kinetic term and the chemical potential do not contribute (since they are delta functional local in $ \S(\t_1, \t_2)$ \cite{Sachdev:2015efa, Gu:2019jub}) and the saddle-point solution is given by
	\begin{align}
	    G^\IR_{\b = \inf}(\t_1, \t_2) \, = \, - \, \sgn(\t_{12}) \, \frac{b^\D \, e^{\sgn(\t_{12}) \pi \mathcal{E}}}{|\t_{12}|^{2\D}} \, ,
  	\label{G_IR}
  	\end{align}
at zero temperature.
In this case, the coefficient is defined by
	\begin{align}
	    b \, = \, \left( \frac{1-2\D}{4\pi}\right) \, \frac{\sin(2\pi \D)}{\cos\pi(\D+i \mathcal{E}) \cos\pi(\D-i \mathcal{E})} \, ,
  	\end{align}
and the ``spectral asymmetry'' $\mathcal{E}$ is related to the chemical potential and it cannot be determined in the IR limit,
but it requires interpolation to the UV solution \cite{Sachdev:2015efa, Davison:2016ngz, Gu:2019jub}.

\subsection{Bulk interpretation of $G_\s$ and $\S_\s$}
\label{sec:bulk interpretation}
In this subsection, we consider a bulk interpretation of the local fields $\{G_\s , \S_\s\}$.
Before considering these fields, let us first remind ourselves about the scalar field coupled to JT gravity in the (near) AdS$_2$
\cite{Almheiri:2014cka, Jensen:2016pah, Maldacena:2016upp, Engelsoy:2016xyb}.
The bulk scalar field is described by the action 
	\begin{align}
		S[\c] \, = \, \frac{1}{2} \int d^2 x \sqrt{g} \, \Big( g^{\m\n} \pa_\m \c \pa_\n \c \, + \, m^2 \c^2 \Big) \, , 
	\end{align}
on the Poincare coordinates
	\begin{align}
		ds^2 \, = \, \frac{d\t^2 + dz^2}{z^2} \, .
	\end{align}
The boundary action is given by
	\begin{align}
		S[\c] \, = \, - \, \frac{1}{2} \int d\t \, \c \pa_z \c \, .
	\label{S_bdy}
	\end{align}
The bulk field is now expressed in terms of the bulk-to-boundary propagator
	\begin{align}
		K_\D(\t, z; \t') \, = \, C_\D \left( \frac{z}{z^2 + (\t -\t')^2} \right)^\D \, ,
	\end{align}
\footnote{where $C_\D = \frac{\G(\D)}{\pi^{\frac{1}{2}}\G(\D -\frac{1}{2})}$,} 
with $\D = 1/2 + \sqrt{m^2 + 1/4}$, as
	\begin{align}
		\c(\t, z) \, = \, \int d\t' \, K_\D(\t, z; \t') j(\t') \, ,
	\end{align}
where the boundary source is defined by
	\begin{align}
		j(\t) \, = \, \lim_{z \to 0} z^{\D - 1} \c(\t, z)\, .
	\end{align}
Using this expression and evaluating the boundary action (\ref{S_bdy}), one finds
	\begin{align}
		S[j] \, = \, - \, (\D -\tfrac{1}{2}) C_\D \int d\t_1 d\t_2 \, \frac{j(\t_1) j(\t_2)}{|\t_1 - \t_2|^{2\D}} \, . 
	\end{align}
If we consider the conformal transformation of the boundary $\t \to f(\t)$, we also have \cite{Almheiri:2014cka, Jensen:2016pah, Maldacena:2016upp, Engelsoy:2016xyb}
	\begin{align}
		S[j] \, = \, - \, (\D -\tfrac{1}{2})C_\D \int d\t_1 d\t_2 \, \frac{|f'(\t_1) f'(\t_2)|^\D}{|f(\t_1) - f(\t_2)|^{2\D}} \, j(\t_1) j(\t_2) \, . 
    \label{S[j]}
	\end{align}

Now we will see that the same effective action is obtained for the fluctuation of the $\S_\s$ field in our partially disorder-averaged SYK model.
\footnote{The discussion below is rather formal in that the mass of a bulk scalar field with $\Delta =1/q$ seems to violate the BF bound.}
To see this, we consider $1/N$ fluctuations of the fields around the totally disorder-averaged IR saddle-point solutions:
	\begin{align}
	    G(\t_1, \t_2) \, &= \, G_{\b=\inf}^{\IR}(\t_1, \t_2) \, + \, \frac{1}{\sqrt{N}} \, \d G(\t_1, \t_2) \, , \nn\\
	    \S(\t_1, \t_2) \, &= \, \S_{\b=\inf}^{\IR}(\t_1, \t_2) \, + \, \frac{1}{\sqrt{N}} \, \d \S(\t_1, \t_2) \, , \nn\\
	    G_\s(\t) \, &= \, G_\s^{\IR}(\t) \, + \, \frac{1}{\sqrt{N}} \, \d G_\s(\t) \, , \nn\\
	    \S_\s(\t) \, &= \, \S_\s^{\IR}(\t) \, + \, \frac{1}{\sqrt{N}} \, \d \S_\s(\t) \, ,
	\end{align}
where $G_\s^{\IR}=\S_\s^{\IR}=0$.
Substituting these expansions into the action (\ref{S_eff}), this generates several interactions among the fluctuations and the background.
The term we are interested in right now is the interaction between the background of the bi-local fields and the fluctuations of the local fields.
This term is given by
	\begin{align}
		S_{\rm int}\big[ G_\s^{\IR}, \d \S_\s \big] \, = \, - \, \frac{1}{4} \int d\t_1 d\t_2 \, \d\S_\s(\t_1) G_{\b=\inf}^{\IR}(\t_1, \t_2) \d\S_\s(\t_2) \, .
	\end{align}
Using the explicit form of the background solution (\ref{GIR_b=inf}), we find
	\begin{align}
		S_{\rm int}\big[ G_\s^{\IR}, \d \S_\s \big] \, = \, - \, \frac{b^\D}{4J^{2\D}} \int d\t_1 d\t_2 \, \frac{\d\S_\s(\t_1) \d\S_\s(\t_2)}{|\t_1 - \t_2|^{2\D}} \, .
	\end{align}
The leading interaction term between the Schwarzian mode and the local field $\d\S_\s$ is given by simply transforming $G^\IR \to G^\IR_f$ (\ref{G^IR_f}) as
	\begin{align}
		S_{\rm int}[f, \d\S_\s] \, = \, - \, \frac{b^\D}{4J^{2\D}} \int d\t_1 d\t_2 \, \frac{|f'(\t_1) f'(\t_2)|^\D}{|f(\t_1) - f(\t_2)|^{2\D}} \, \d\S_\s(\t_1) \d\S_\s(\t_2) \, .
	\end{align}
This form of the boundary action precisely agrees with (\ref{S[j]}).
This implies that the bulk interpretation of the local field $\d\S_\s$ is a boundary source of extra bulk fields.
Then, it is natural to interpret the dual field $\d G_\s$ as the expectation value of the corresponding boundary operator.

\subsection{Comments on general $\s$ case}
\label{sec:general sigma}
In this subsection, we make some comments on the general $\s$ case.
As we mentioned below (\ref{inverse}), the time translation symmetry of $\S(\t_1, \t_2)$ requires from (\ref{delta G}) that 
the only possible solution for the local field $\S_\s$ is a time-independent constant solution.
Then, as we explained in section \ref{sec:slightly fixed-coupling},
such a constant term in (\ref{delta G}) has the same structure as the chemical potential contribution in the complex SYK model,
and the solution of $G(\t_1, \t_2)$ is given by (\ref{G_free}) for free case or by (\ref{G_IR}) for the IR case.

Given these facts, we can evaluate the integral in (\ref{delta Sigma_sigma}) for example for $ G_{\rm free}$ as
	\begin{align}
		\int_0^\b d\t' \, G_{\rm free}(\t, \t') \, = \, \frac{1-e^{\m \t}}{\m(1+e^{\b \m})} \, + \, \frac{e^{\m(\b - \t)}-1}{\m(1+e^{-\b \m})} \, .
	\end{align}
This seems to indicate that we can still have a nontrivial time-dependent solution for $G_\s(\t)$ even though we should have a time-independent constant solution for $\S_\s$.
At the same time, we also have to make the solutions consistent with the other equation (\ref{delta G_sigma}).
This seems quite nontrivial for the general $\s$ case.
It would be very interesting to study this general $\s$ case more in detail, but we leave this to future work.

\section{Two-point Function $\big\la Z_L Z_R\big\ra_J$}
\label{sec:2pt function}
In this section, we study the partially disorder-averaged two-point partition functions in the large $N$ limit.
Before the partial disorder averaging, we have the two-point function of the partition functions
	\begin{align}
		Z_L Z_R \, = \, \int D\c_i^L D\c_i^R \,
		\exp\bigg[ &- \frac{1}{2} \sum_{a=L,R} \int d\t \, \sum_{i=1}^N \c_i^a \pa_\t \c_i^a \nn\\
		&\qquad + \, i^{\frac{q}{2}} \sum_{a=L,R} \int d\t \, \sum_{i_1<\cdots<i_q}^N J_{i_1 \cdots i_q} \, \c_{i_1}^a \cdots \c_{i_q}^a \bigg] \, .
 	\label{Z_LZ_R}
 	\end{align}
Using the duplicated Hubbard–Stratonovich transformation
	\begin{align}
		1 \, &= \, \int \prod_{a,b}^{L,R} DG_{ab} \, \d\left( G_{ab} - \frac{1}{N} \sum_{i=1}^N \c_i^a \c_i^b \right) \int \prod_a^{L,R} DG_\s^a \, \d\left( G_\s^a - \frac{1}{N} \sum_{i=1}^N \th_i \c_i^a \right) \nn\\
		&= \, \int \prod_{a,b}^{L,R} DG_{ab} D\S_{ab} \exp\left[  - \frac{1}{2} \Tr \, \S_{ab} \cdot \left( N G_{ab} - \sum_{i=1}^N \c_i^a \c_i^b \right) \right] \nn\\
		&\qquad \quad \times \int \prod_a^{L,R} DG_\s^a D\S_\s^a \exp\left[ - \Tr \, \S_\s^a \cdot \left( N G_\s^a - \sum_{i=1}^N \th_i \c_i^a \right) \right] \, ,
 	\end{align}
we now find 
	\begin{align}
		\big\la Z_L Z_R \big\ra_J \, = \, \int \prod_{a,b}^{L,R} DG_{ab} D\S_{ab} \int \prod_a^{L,R} DG_\s^a D\S_\s^a \ e^{-S_{\rm eff}[G_{ab}, \S_{ab}, G_\s^a, \S_\s^a]} \, ,
 	\end{align}
with
	\begin{align}
		S_{\rm eff}[G_{ab}, \S_{ab}, G_\s^a, \S_\s^a] \, &= \, - \frac{N}{2} \sum_{a,b} \Tr\log (\S_{ab})
		- \frac{N}{2} \sum_a \int d\t \big[ \pa_\t G_{aa}(\t, \t')\big]_{\t'=\t} \nn\\
		&\quad + \frac{N}{2} \sum_{a,b} \int d\t_1 d\t_2 \left( \S_{ab}(\t_1, \t_2) G_{ab}(\t_1, \t_2) - \frac{\wtd{J}^2}{q} G_{ab}(\t_1, \t_2)^q \right) \nn\\
		&\quad + N \sum_a \int d\t \left( \S_\s^a(\t) G_\s^a(\t) - \frac{J_\s}{q} G_\s^a(\t)^q \right) \nn\\
		&\quad - \, \frac{N}{4} \sum_{a,b} \int d\t_1 d\t_2 \, \S_\s^a(\t_1) G_{ab}(\t_1, \t_2) \S_\s^b(\t_2) \, .
	\label{S_eff2}
	\end{align}

Similar to (\ref{delta G}) - (\ref{delta Sigma_sigma}), the large $N$ saddle-point equations are
	\begin{align}
		\d G_{ab} \, :& \qquad \S_{ab}(\t_1, \t_2) \, = \, \wtd{J}^2 \, G_{ab}(\t_1, \t_2)^{q-1} + \d_{ab} \, \pa_{\t_1} \d(\t_1 - \t_2)
		\, + \, \frac{1}{2} \S_\s^a(\t_1) \S_\s^b(\t_2)\, , \nn \\
		\d \S_{ab} \, :& \qquad \S_{ab}^{-1}(\t_1, \t_2) \, = \, G_{ab}(\t_1, \t_2) \, ,\nn \\
		\d G_\s^a \, :& \qquad \S_\s^a(\t) \, = \, J_\s \, G_\s^a(\t)^{q-1} \, ,\nn \\
		\d \S_\s^a \, :& \qquad G_\s^a(\t) \, = \, \frac{1}{4} \sum_b \int d\t' \, G_{ab}(\t, \t') \S_\s^b(\t') \, . \label{saddlepointeq}
  	\end{align}

\subsection{Zero-dimensional SYK}
\label{sec:0d SYK2}

\subsubsection{Saddle-point solutions}
The large $N$ saddle-point equations are a little complicated to solve, so let us first consider the zero-dimensional case again.
In this case, we only have $\{ G_{LR}, \S_{LR} \}$ for the bi-local fields.
For the following discussion, it is more convenient to combine the two couplings into one effective coupling that appears in the interaction term between the local and bi-local sectors. 
This can be implemented by rescaling the Hubbard–Stratonovich fields as $G_{LR} \to \wtd{J}^{-2/q} G_{LR}$, $\S_{LR} \to \wtd{J}^{2/q} \S_{LR}$,
$G_\s \to J_\s^{-1/q} G_\s$ and $\S_\s \to J_\s^{1/q} \S_\s$.
After all of these rescalings, the effective action of this case reads
	\begin{align} 
		S_{\rm eff}[G_{LR}, \S_{LR}, G_\s^a, \S_\s^a] \, &= \, - N \log (\S_{LR}) \, + \, N \left( \S_{LR} G_{LR} - \frac{1}{q} \, G_{LR}^q \right)  \nn \\
		&\quad +N \sum_a \left( \S_\s^a G_\s^a - \frac{1}{q} \big( G_\s^a \big)^q \right) \, - \, N \lambda_\sigma G_{LR} \S_\s^L \S_\s^R\, ,\label{S_eff-0d}
	\end{align}
where we introduced the coupling between the local and bi-local sectors by
    \begin{align}
        \lambda_\s \, \equiv \, \frac{1}{2} \left( \frac{J_\s}{\wtd{J}} \right)^{\frac{2}{q}} \, .
    \end{align}
This coupling behaves $\lambda_\sigma\sim \frac{1}{2}(J_0/\sigma)^{2/q}$ as $\sigma\rightarrow 0$ i.e., in the totally fixed-coupling limit and $\lambda_\sigma\sim \frac{1}{2}(JJ_0/\sigma^2)^{2/q}$ as $\sigma\rightarrow \infty$ i.e., in the total disorder-averaged limit.    
Here from the action, we excluded the constant contribution $-\frac{2N}{q} \log \wtd{J}$ coming from the rescaling of $\S_{LR}$,
so that the two-point function of the partition functions is now written as
	\begin{align}
		\big\la Z_L Z_R \big\ra_J \, = \, \wtd{J}^{\, \frac{2N}{q}} \int dG_{LR} d\S_{LR} \int \prod_a^{L,R} dG_\s^a d\S_\s^a \
		e^{-S_{\rm eff}[G_{LR}, \S_{LR}, G_\s^a, \S_\s^a]} \, .
 	\label{<Z_LZ_R>}
 	\end{align}

Now the saddle-point equations are written as
	\begin{gather}
		\S_{LR} \, = \, \big( G_{LR} \big)^{q-1} + \, \l_\s \, \S_\s^L \S_\s^R\, , \qquad 
		\frac{1}{\S_{LR}} \, = \, G_{LR} \, , \label{G^LR-eq} \\[2pt]
		\S_\s^a \, = \,  \,  \, \big( G_\s^a \big)^{q-1} \, , \qquad
		G_\s^L \, = \, \l_\s \, G_{LR} \S_\s^R \, , \qquad 
		G_\s^R \, = \, \l_\s \, G_{LR} \S_\s^L \, . \label{Sigma_sigma-eq}
  	\end{gather}
Let us first solve for $\{G_\s , \S_\s\}$ in terms of $G_{LR}$. From (\ref{Sigma_sigma-eq}), we can find one trivial solution and $q(q-2)$ nontrivial solutions
\footnote{These $q(q-2)$ solutions degenerate into $q$ independent solutions after plugging the solutions of $G_{LR}$.}
	\begin{align}
	    \big\{ G_\s^L \, , G_\s^R \big\} \, = \, \{0, 0\} \, , \quad
		\left\{	e^{\frac{2n\pi i}{q(q-2)}} \big( \lambda_\sigma G_{LR} \big)^{-\frac{1}{q-2}} , \,
        e^{\frac{(q-1)2n\pi i}{q(q-2)}} \big( \lambda_\sigma G_{LR} \big)^{-\frac{1}{q-2}} \right\} \, ,
	\label{WH&HW}
  	\end{align}
with $n=1,2, \cdots, q(q-2)$. The corresponding solutions of $\S_\s^a$ are given by the first equation of (\ref{Sigma_sigma-eq}).
The nontrivial solutions are valid for $q>2$. In the rest of this subsection, we focus on $q>2$ case, and we will present a discussion for $q=2$ in Appendix~\ref{app:q=2}.
We will call the first trivial solution $\{ G_\s^L \, , G_\s^R \} = \{0, 0\}$ ``wormhole'' saddle-point solution,
while the other $q(q-2)$ nontrivial solutions are {\it analogous} to the ``half-wormhole'' saddle-point solutions.
\footnote{Even though in the main text, we often call these nontrivial solutions ``half-wormhole'' solutions,
we emphasize that these solutions are {\it analogous} to the half-wormhole saddle-point solutions
found in \cite{Saad:2021rcu} and discussed further in \cite{ Mukhametzhanov:2021nea}.
There are some differences between our nontrivial solutions and their half-wormhole saddle-point solutions, as we will discuss below and in section~\ref{sec:conclusions}.
}
The reasoning for these names will be explained below.

\begin{figure}[t!]
 \begin{center}
\includegraphics[scale=0.2]{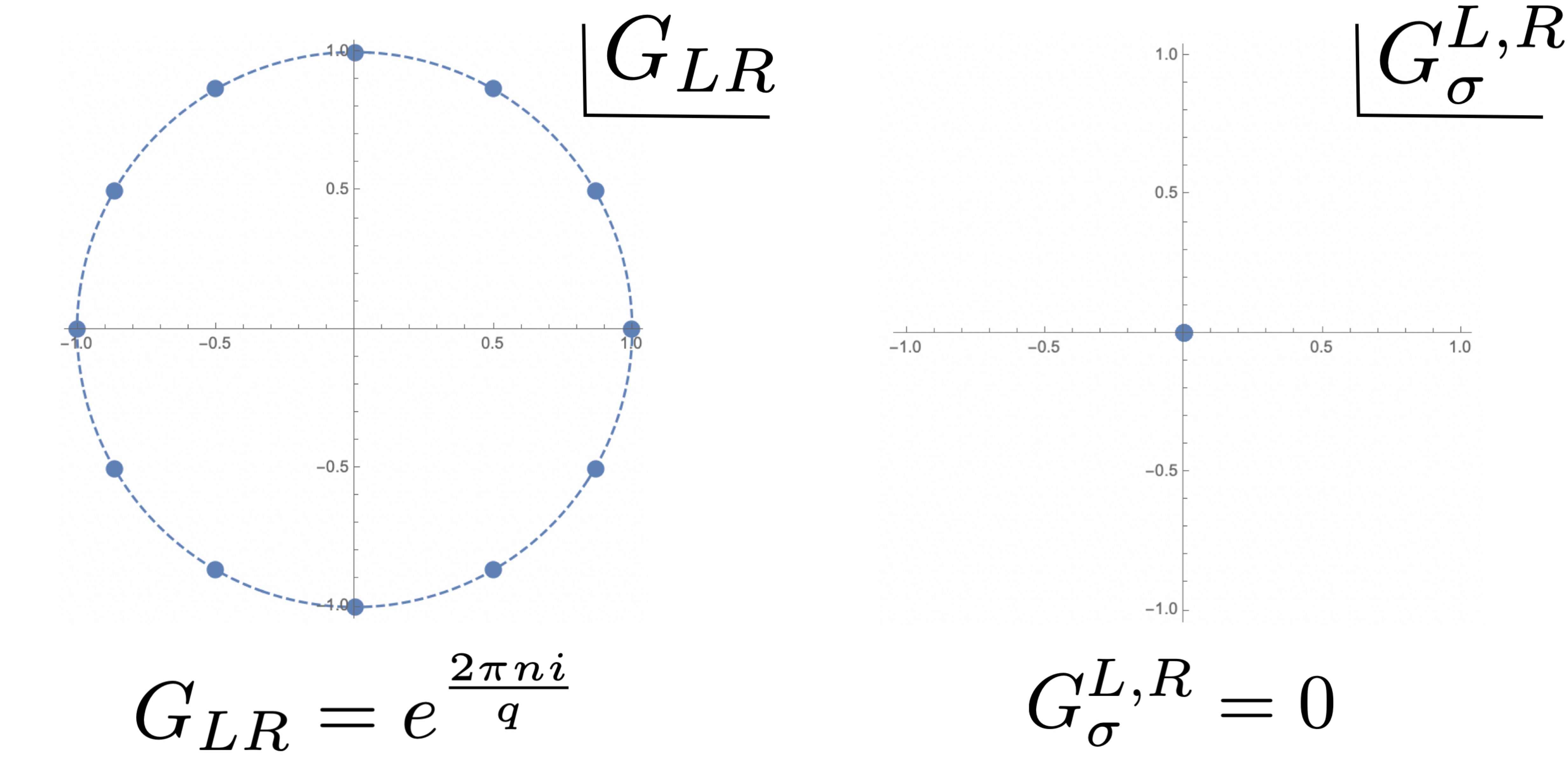}
\end{center}
 \caption{The wormhole saddles for $q=12$.}
\label{fig:saddles1}
\end{figure}

For the wormhole saddles $\{ G_\s^L \, , G_\s^R \} = \{0, 0\}$, the remaining equations (\ref{G^LR-eq}) are solved as
	\begin{align}
		\qquad \big\{ G_{LR} \, , \S_{LR} \big\} \, = \, \left\{ \, e^{\frac{2m\pi i}{q}} \, , \, e^{-\frac{2m\pi i}{q}} \, \right\} \, , \qquad
		{\rm with} \qquad m \, = \, 1, 2, \cdots, q \, .
    \label{wormhole}
  	\end{align}
There are $q$ solutions and they are located at $|G^{LR}|=1$ and $|\S^{LR}|=1$, which agree with the wormhole saddles found in \cite{Saad:2021rcu}.
This is why we call the trivial solution $\{ G_\s^L \, , G_\s^R \} = \{0, 0\}$ the wormhole saddle-point solution.
These wormhole solutions are depicted in Figure~\ref{fig:saddles1}.

\begin{figure}[t!]
 \begin{center}
\includegraphics[scale=0.2]{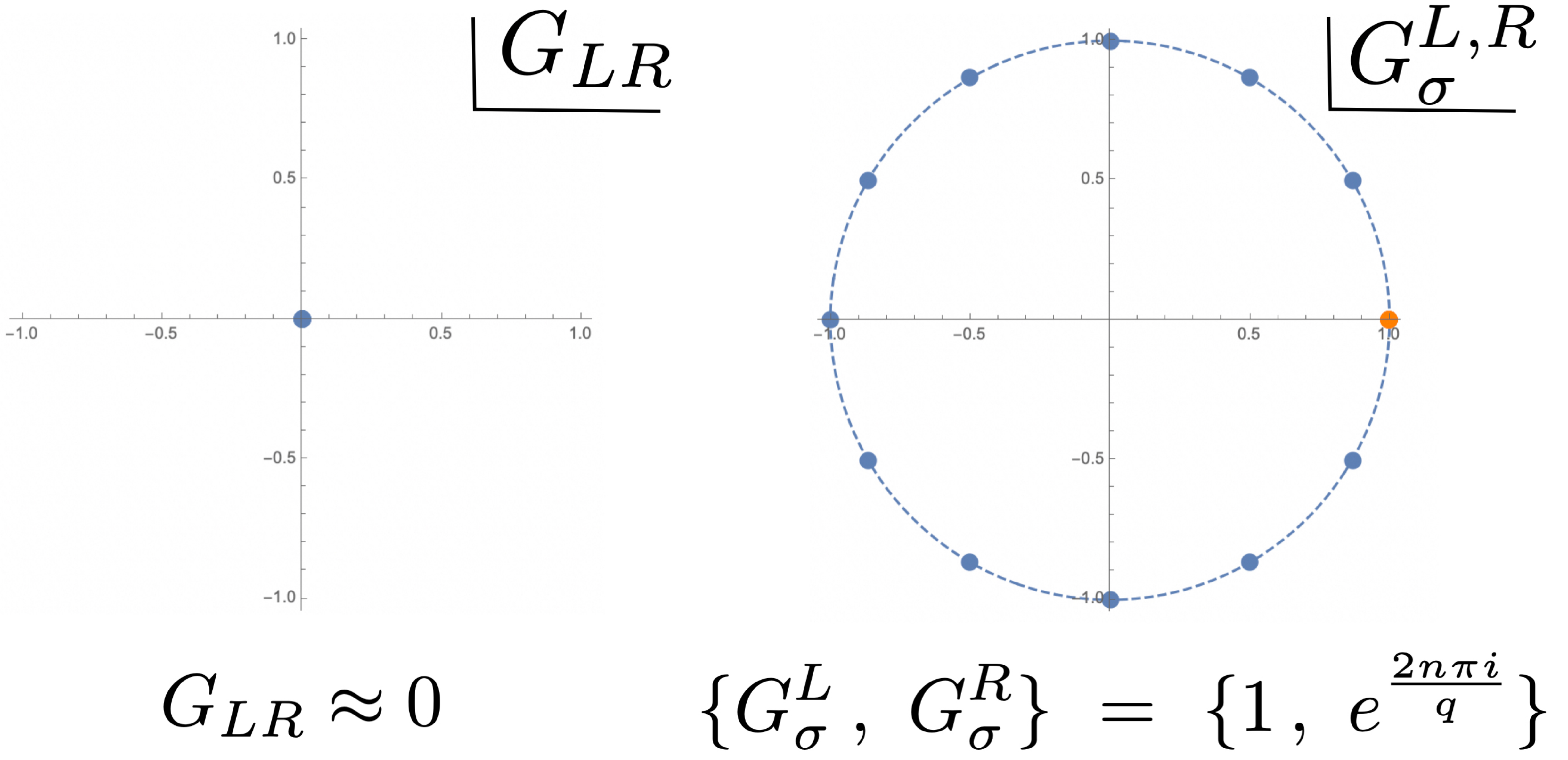}
\end{center}
 \caption{The half-wormhole saddles in the totally fixed-coupling limit for $q=12$.}
\label{fig:saddles2}
\end{figure}

For the half-wormhole saddles, we look for solutions that behave differently from the above wormhole solutions in the totally fixed-coupling limit ($\s \to 0$).
Hence, in the $\s \to 0$ limit for the first equation, we look for solutions satisfy $\S_{LR} \approx \lambda_\s \S_\s^L \S_\s^R$. This is solved as
	\begin{align}
    	\left\{ G_{LR}\, ,	\S_{LR}\right\} \, \approx \,\left\{ \lambda_\sigma^{-1} \, e^{\frac{2n\pi i}{q}} \, , \, \lambda_\sigma \, e^{\frac{-2n\pi i}{q}} \right\} \, ,  
  	\label{Sigma_LR2}
  	\end{align}
and 
  	\begin{align}
	    \left\{\big\{ G_\s^L \, , G_\s^R \big\} \, ,\ \big\{ \Sigma_\s^L \, , \Sigma_\s^R \big\}\right\} \ \approx& \ \left\{
		\left\{ 1,\, e^{\frac{2n\pi i}{q}} \right\}\, ,\ \left\{ 1,\, e^{\frac{-2n\pi i}{q}} \right\} \right\}\nonumber\\  \ {\rm or}& \ \left\{ \left\{ e^{\frac{2n\pi i}{q}} , \, 1 \right\}\, , \  \left\{ e^{\frac{-2n\pi i}{q}}\, , 1 \right\}\right\} \, .
  	\end{align}
Notice that we have $2q$ independent solutions similarly to the wormhole solutions.
Since in the totally fixed-coupling limit ($\s \to 0$), we have $\lambda_\sigma^{-1} \propto \s^{-2/q} \to 0$, we have $G_{LR} \to 0$ in this limit ($\s \to 0$).
In contrast to the wormhole solutions, this means that we have {\it no} correlation between the two SYK systems (L and R)
and they are analogous to the half-wormhole saddle-point solutions found in \cite{Saad:2021rcu} and discussed further in \cite{Mukhametzhanov:2021nea}.
These half-wormhole solutions are depicted in Figure~\ref{fig:saddles2}.

On the other hand, in the disorder-averaged limit ($\s \to \inf$), the coupling $\lambda_\sigma$ becomes small, and $G_{LR}$ becomes large.
Therefore, the approximation $\S_{LR} \approx \lambda_\s \S_\s^L \S_\s^R$ we used is no longer valid in this limit,
and we should treat as the extra term $\lambda_\s \S_\s^L \S_\s^R$ as a perturbation around the wormhole solutions.

\subsubsection{On-shell action}
In this subsection, we study the $\sigma$-dependence of the  on-shell action of the saddle-point solutions found above.

Let us first consider the on-shell action by taking a derivative with respect to explicit $\l_\sigma$ dependence as in \cite{Maldacena:2016hyu}
	\begin{align}
		\frac{\pa}{\pa \lambda_\sigma } \, S_{\rm eff} \, = \, - \, N G_{LR} \S_\s^L \S_\s^R \, .
  	\end{align}
For the wormhole solutions, it is obvious that
	\begin{align}
		\frac{\pa}{\pa \lambda_\sigma} \, S^{\onshell}_{\WH} \, = \, 0 \, , 
  	\end{align}
thus the on-shell action of the wormhole solutions does not have $\l_\sigma$ dependence.
For the half-wormhole solutions (\ref{Sigma_LR2}) for small $\s$, we can write the derivative of the on-shell action as 
	\begin{align}
		\frac{\pa}{\pa \lambda_\sigma} \, S^{\onshell}_{\HW}  \, \approx \, -N \lambda^{-1}_{\sigma} \, .
  	\label{dF_HW}
  	\end{align}

Let us also study the value of the on-shell action directly (without taking derivative with respect to $\l_\s$).
For the wormhole solutions, this is given by
	\begin{align}
		S^{\onshell}_{\WH} \, &= \, - N \log \S_{LR} \big|_{\onshell} \, + \, N \left( 1 - \frac{1}{q} \right) \, G_{LR}^q \big|_{\onshell} \nn\\
		&= \, N \left( 1 - \frac{1}{q} \right) \, + \, \frac{2m\pi i N}{q} \, .
	\end{align}
For the half-wormhole solutions at small $\s$, we have $G_{LR} \approx 0$. Therefore, it is given by
	\begin{align}
		S^{\onshell}_{\HW}  \, &\approx \, \left[ - N \log \S_{LR}
		\, + \, N  \left( 1 - \frac{1}{q} \right) \sum_a \big( G_\s^a \big)^q  \right]_{\onshell} \nn\\
        &\approx \, - \, N \log( \l_\s ) \, + \, 2N \left( 1 - \frac{1}{q} \right) \, + \, \frac{2n\pi i N}{q} \, .
    \label{F_HW}
	\end{align}
We can see that the $\l_\s$ dependence agrees with that obtained by integrating (\ref{dF_HW}).

For $\s \to \inf$, we have only wormhole contributions. Therefore, combining with the prefactor in (\ref{<Z_LZ_R>}), the two-point function is given by
	\begin{align}
		\big\la Z_L Z_R \big\ra_J \, &\approx \, J^{\, \frac{2N}{q}} \sum_{m=1}^q e^{-N(1-\frac{1}{q}) - \frac{2m\pi i N}{q}} \nn\\
		&= \, q \, J^{\, \frac{2N}{q}} \, e^{-N(1-\frac{1}{q})} \, ,
 	\end{align}
where we used $\wtd{J} \to J$ in the $\s \to \inf$ limit, and the summation (\ref{m-sum}).
On the other hand, for $\s \to 0$, we have $\wtd{J} \to 0$. Therefore, the wormhole contributions, which do not have $\s$ dependence, do not contribute.
For the half-wormhole contributions, the $\s$ dependence coming from (\ref{F_HW}) precisely cancels with the $\s$ dependence coming from the prefactor in (\ref{<Z_LZ_R>}) and we find
	\begin{align}
		\big\la Z_L Z_R \big\ra_J \, &\approx \, 2^{-N} J_{0}^{\frac{2 N}{q}} \sum_{n=1}^{q} e^{ - 2N\left(1-\frac{1}{q}\right) - \frac{2 n \pi i N}{q}} \nn\\
		&= \, q \, 2^{-N} J_{0}^{\frac{2 N}{q}} \, e^{ - 2N \left(1-\frac{2}{q}\right)} \, .
 	\label{<Z_LZ_R>-small-sigma}
 	\end{align}
We can see that the $J$ and $J_0$ dependence of the two-point function precisely agree with the results of \cite{Mukhametzhanov:2021nea} in both limits.
Also the above $\s\to 0$ result precisely agrees with the square of (\ref{<Z>_J-exact}) with $\th_i^2 =1/2$ in the $\s\to 0$ limit.
Furthermore, at $\s = 0$, we can compute the two-point function directly (which is presented in Appendix~\ref{app:direct computation})
and the result agrees with the above computation.

The large $N$ saddles contributing to this totally fixed-coupling two-point function look different from the conclusion of \cite{Saad:2021rcu, Mukhametzhanov:2021nea},
but this is simply due to the difference of the ``linked'' half-wormhole and the ``unlinked'' half-wormhole, as we will explain further in section~\ref{sec:conclusions}.

\subsection{Adding a coupling}
\label{sec:adding a coupling}
Let us make some comments what happens if we introduce a coupling between $L$ and $R$ systems in zero-dimensional SYK model
\begin{align}
S_{\rm int}=\mu\sum_{i=1}^N \c_i^L\c_i^R \, ,
\end{align}
according to \cite{Maldacena:2018lmt, Saad:2021rcu}. 
The saddle-point equations become
\begin{gather}
		\S_{LR} \, = \, \big( G_{LR} \big)^{q-1} + \, \l_\s \, \S_\s^L \S_\s^R-\mu\, , \qquad 
		\frac{1}{\S_{LR}} \, = \, G_{LR} \, , \\[2pt]
		\S_\s^a \, = \,  \,  \, \big( G_\s^a \big)^{q-1} \, , \qquad
		G_\s^L \, = \, \l_\s \, G_{LR} \S_\s^R \, , \qquad 
		G_\s^R \, = \, \l_\s \, G_{LR} \S_\s^L \, .
  	\end{gather}
 If we consider the strong coupling limit near the totally fixed coupling region such that $1 \ll\lambda_\sigma \ll \mu$, the interaction term can be neglected. In this case, the half-wormhole saddle and the wormhole saddle degenerates, and we have a solution
 \begin{align}
     G_{LR}\approx\mu^{\frac{1}{q-1}},\ \Sigma_{LR}\approx \mu^{-\frac{1}{q-1}}\, ,
 \end{align}
with $G^{L,R}_{\sigma}\approx 0, \Sigma^{L,R}_{\sigma}\approx 0$.
Therefore, even near the fixed coupling region, the wormhole contribution becomes dominant.
This is consistent with the conclusions \cite{Saad:2021rcu, Mukhametzhanov:2021nea}.

\subsection{Brownian SYK}
\label{sec:brownian}
Next, we study the Brownian SYK case discussed in \cite{Saad:2018bqo}.
From the effective action (\ref{S_eff2}), the Brownian SYK corresponds to reducing the bi-local fields to local fields.
Furthermore, in this reduction, the diagonal fields in the $a$ index reduce to constants due to the anti-commutation relation of the fermion (\ref{ACR}), so we only need to keep $\{G_{LR} , \S_{LR}\}$.
Therefore, the effective action is now given by 
	\begin{align}
		&S_{\rm eff}[G_{LR}, \S_{LR}, G_\s^a, \S_\s^a] \, = \, - \frac{N}{2} \Tr \log ( \S_{LR})
		+ \frac{N}{2} \int d\t \left( \S_{LR}(\t) G_{LR}(\t) - \frac{\wtd{J}^2}{q} G_{LR}(\t)^q \right) \nn\\
		&\qquad + N \sum_a \int d\t \left( \S_\s^a(\t) G_\s^a(\t) - \frac{J_\s}{q} G_\s^a(\t)^q \right) 
		\, - \, \frac{N}{4} \int d\t \, G_{LR}(\t) \S_\s^L(\t) \S_\s^R(\t) \, .
	\end{align}

In general, it is still hard to search for time-dependent saddle-point solutions.
Hence, let us now consider static (time-independent) saddle-point solutions.
As explained in \cite{Saad:2018bqo}, in this case, we have
	\begin{align}
		\frac{1}{2} \Tr \log ( \S_{LR}) \, \to \, \log\left( 2 \cosh\left( \frac{\b \S_{LR}}{4} \right) \right) \, \approx \, \pm \frac{\b \S_{LR}}{4} \, , 
    \label{brownian-large beta}
	\end{align}
where for the rightmost equation, we take the low-temperature limit (i.e., large $\b$) and the plus/minus sign depending on the real part of $\S_{LR}$.
Then, the large $N$ saddle-point equations are
	\begin{gather}
		\S_{LR} \, = \, \wtd{J}^2 \, G_{LR}^{q-1} \, + \, \frac{1}{2} \S_\s^L \S_\s^R\, , \qquad 
		\pm \frac{1}{2} \, = \, G_{LR} \, , \label{G^LR-eq2} \\[2pt]
		\S_\s^a \, = \, J_\s \, \big( G_\s^a \big)^{q-1} \, , \qquad
		G_\s^L \, = \, \frac{1}{4} \, G_{LR} \S_\s^R \, , \qquad 
		G_\s^R \, = \, \frac{1}{4} \, G_{LR} \S_\s^L \, , \label{Sigma_sigma-eq2}
  	\end{gather}
We can easily see that the equations for $\{G_\s^a ,\S_\s^a \}$ (\ref{Sigma_sigma-eq2}) are structurally identical to the 0d SYK case (\ref{Sigma_sigma-eq}).
Therefore, the wormhole and half-wormhole solutions are given by 
	\begin{align}
	    \big\{ G_\s^L \, , G_\s^R \big\} \, = \, \{0, 0\} \, , \quad
		\left\{	e^{\frac{2n\pi i}{q(q-2)}} \left( \frac{J_\s}{8} \right)^{-\frac{1}{q-2}} , \,
        e^{\frac{(q-1)2n\pi i}{q(q-2)}} \left( \frac{J_\s}{8} \right)^{-\frac{1}{q-2}} \right\} \, ,
  	\end{align}
where we also used $G_{LR} =\pm 1/2$.

For the wormhole solution $\{ G_\s^L \, , G_\s^R \} = \{0, 0\}$, the remaining equations (\ref{G^LR-eq2}) are solved as
	\begin{align}
		\qquad \big\{ G_{LR} \, , \S_{LR} \big\} \, = \, \bigg\{ \, \pm \frac{1}{2} \ , \ \pm \frac{\wtd{J}^2}{2^{q-1}} \, \bigg\} \, .
  	\end{align}
These wormhole solutions agree with the ones found in \cite{Saad:2018bqo}.
\footnote{The imaginary factor $i$ difference is due to the fact that we study the model in Euclidean signature while \cite{Saad:2018bqo} used Lorentzian time.}
We emphasize that the wormhole solutions are obtainable {\it without} continuing to Lorentzian time, but simply taking the low-temperature limit in Euclidean time.
For the half-wormhole solutions, the $\S_{LR}$ solution is obtained as
	\begin{align}
		\S_{LR} \, = \, \pm \frac{\wtd{J}^2}{2^{q-1}} \, + \, \frac{1}{2} \left( \frac{J_\s}{8} \right)^{-\frac{2}{q-2}} e^{\frac{2n\pi i}{q-2}} \, .
  	\end{align}

The on-shell actions for these solutions are computed as
	\begin{align}
		S^{\onshell}_{\WH} \, &= \, \mp \, \frac{N\b}{4} \S_{LR} \big|_{\onshell}
		\, + \, \frac{N\b \wtd{J}^2}{2} \left( 1 - \frac{1}{q} \right) \, G_{LR}^q \big|_{\onshell} \nn\\
		&= \, - \, \frac{N \b \wtd{J}^2}{q \, 2^{q+1}} \, ,
	\end{align}
for the wormhole solutions and
	\begin{align}
		S^{\onshell}_{\HW}  \, &= \, \left[ \mp \, \frac{N \b}{4} \S_{LR} + \frac{N \b \wtd{J}^2}{2} \left( 1 - \frac{1}{q} \right) G_{LR}^q
		+ N \b J_\s \left( 1 - \frac{1}{q} \right) \sum_a \big( G_\s^a \big)^q  \right]_{\onshell} \nn\\
        &= \, - \, \frac{N \b \wtd{J}^2}{q \, 2^{q+1}} \, \mp \, \frac{N \b}{8} \left( \frac{J_\s}{8} \right)^{-\frac{2}{q-2}}
        \, + \, 8N \b \left( 1 - \frac{1}{q} \right) \left( \frac{J_\s}{8} \right)^{-\frac{2}{q-2}} \Big( e^{\frac{2n \pi i}{q-2}} + e^{\frac{(q-1)2n \pi i}{q-2}} \Big) \, .
	\end{align}
for the half-wormhole solutions.

For the total disorder-average limit ($\s \to \inf$), in order to keep the large $\b$ approximation (\ref{brownian-large beta}) valid, we need $\b J^2$ large. 
Therefore,both the wormhole and the half-wormhole solutions contribute equally to the disorder-averaged two-point function  in this limit.
This behavior is different from the one found in the zero-dimensional case in section~\ref{sec:0d SYK2}.
Again, we expect that this difference is associated with the difference of the ``linked'' and ``unlinked'' half-wormhole, as we will explain further in section~\ref{sec:conclusions}.

\section{Spectrum}
\label{sec:spectrum}
In this section, we study the spectrum of the partially disorder-averaged SYK model.
To see the spectrum, it is more convenient to incorporate the $\sigma$-dependence into the Hamiltonian itself,
such that the probability distribution of the newly defined random coupling has the original distribution form.
The $\sigma$-dependence in the probability distribution can be eliminated as
\begin{align}
\mathcal{N}_\s^{-1} P(J_{i_1 \cdots i_q}) \, &= \, \mathcal{N}_\s^{-1} \exp\left[ -\frac{N^{q-1}}{2(q-1)!} \sum_{i_1<\cdots<i_q}^N
		\left( \frac{J_{i_1 \cdots i_q}^2}{J^2} \, + \, \frac{(J_{i_1 \cdots i_q} - J_{i_1 \cdots i_q}^{(0)})^2}{\s^2} \right)\right]\nonumber\\
		&= \, \left( \frac{N^{q-1}}{2(q-1)! \pi \wtd{J}^2} \right)^{\frac{1}{2} \binom{N}{q}} \exp\left[ -\frac{N^{q-1}}{2(q-1)!} \sum_{i_1<\cdots<i_q}^N
		 \frac{\hat{J}^2_{i_1 \cdots i_q}}{J^2}\right]
		\, ,
\end{align}
by introducing a new coupling constant 
\begin{align}
    \hat{J}_{i_1 \cdots i_q}\equiv \sqrt{1+\frac{J^2}{\sigma^2}}\left[J_{i_1 \cdots i_q}-\frac{1}{\left(1+\frac{\sigma^2}{J^2}\right)}J^{(0)}_{i_1 \cdots i_q}\right]\, .
\end{align}
The Hamiltonian of the SYK model can be written in terms of $ \hat{J}_{i_1 \cdots i_q}$ as
\footnote{A similar (but different) deformation of the SYK model is studied in the context of $T\bar{T}$ deformation in \cite{Gross:2019uxi}.}
\begin{align}
\label{pfcSYK}
H \, = \, \frac{i^{\frac{q}{2}}}{\sqrt{1+\frac{J^2}{\sigma^2}}} \sum_{i_1<i_2<\cdots<i_q}^N \hat{J}_{i_1 \cdots i_q} \, \c_{i_1} \cdots \c_{i_q}
\, + \, \frac{i^{\frac{q}{2}}}{\left(1+\frac{\sigma^2}{J^2}\right)}\sum_{i_1<i_2<\cdots<i_q}^N J^{(0)}_{i_1 \cdots i_q} \, \c_{i_1} \cdots \c_{i_q}\, .
\end{align}
This Hamiltonian manifests the transition between the usual disorder-averaged SYK model and the fixed-coupled SYK model, i.e., 
\begin{align}\label{Hlimit}
H \, &\underset{\sigma \rightarrow \infty}{\rightarrow} \, H_{\hat{J}}\equiv i^{\frac{q}{2}}\sum_{i_1<i_2<\cdots<i_q}^N \hat{J}_{i_1 \cdots i_q} \, \c_{i_1} \cdots \c_{i_q}\, ,\nonumber\\
H \, &\underset{\sigma \rightarrow 0}{\rightarrow} \, H_{J^{(0)}}\equiv i^{\frac{q}{2}}\sum_{i_1<i_2<\cdots<i_q}^N J^{(0)}_{i_1 \cdots i_q} \, \c_{i_1} \cdots \c_{i_q}\, .
\end{align}
In  the $\sigma\rightarrow 0$ limit, the Hamiltonian becomes independent of $\hat{J}_{i_1 \cdots i_q}$, thus the average over the random coupling is performed trivially as
\begin{align}
\mathcal{N}_\s^{-1} \int \prod_{i_1<\cdots<i_q}^N dJ_{i_1 \cdots i_q} \, P(J_{i_1 \cdots i_q})=1\, .
\end{align}

\begin{figure}[t!]
 \begin{center}
\includegraphics[scale=0.7]{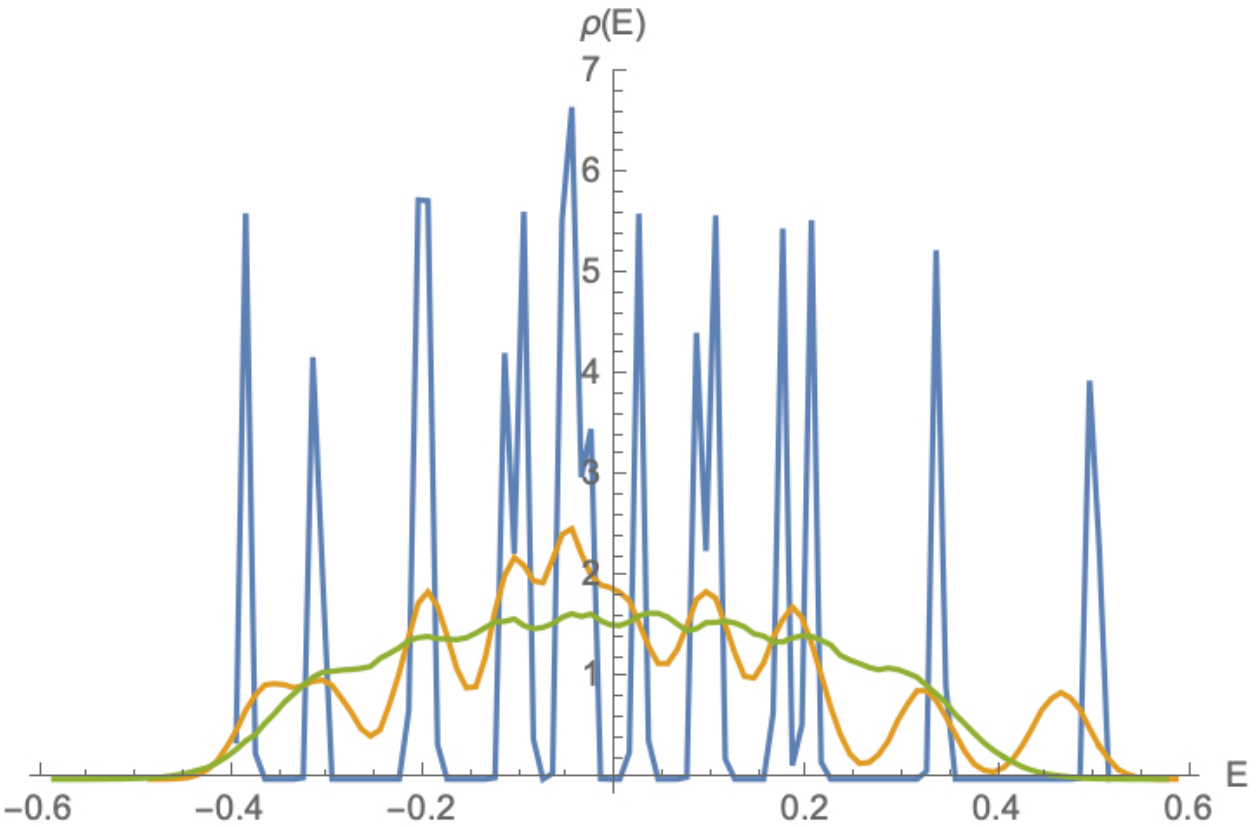}
\end{center}
 \caption{Density of states $\rho(E)$ with  $\sigma=0.03$(blue), $0.3$(orange), $3$(green). We take $N=8$, $q=4$ and $J=1$, and take $50000$ samples for averaging. In this plot, the external coupling $J^{(0)}_{i_1 \cdots i_q}$ is chosen randomly from the Gaussian probability distribution $P(J^{(0)}_{i_1 \cdots i_q}) \, = \, \exp\left(
 -\frac{N^{q-1}}{2(q-1)!J^2} \sum_{i_1<\cdots<i_q}^N
	 (J^{(0)}_{i_1 \cdots i_q})^2 \right)$ with $N=8,q=4$ and $J=1$.}
\label{fig:Spectrum}
\end{figure}

The partition function $\langle Z(it)\rangle_J$ of the partially disorder-averaged SYK model in the $\sigma\rightarrow 0$ limit simply becomes that in the SYK model with a fixed coupling $J^{(0)}_{i_1 \cdots i_q} $, i.e.,
\begin{align}
\langle Z(it)\rangle_J=\sum_{E_{J^{(0)}}}{\rm exp}\left[-it E_{J^{(0)}} \right]\, ,
\end{align}
where $E_{J^{(0)}}$ are the eigenvalues of the Hamiltonian $H_{J^{(0)}}$. Then the density of states, which is given by the Fourier transform of $\langle Z(it)\rangle_J$, becomes a sum of the delta functions
\begin{align}
\langle\rho(E)\rangle_J &=\int^\infty_{-\infty} \frac{dt}{2\pi}e^{-it E} \langle Z(it)\rangle_J\nonumber\\
&=\sum_{E_{J^{(0)}}}\int^\infty_{-\infty} \frac{dt}{2\pi}e^{-it (E-E_{J^{(0)}})}=\sum_{E_{J^{(0)}}}\delta(E-E_{J^{(0)}})\, .
\end{align}
In Figure~\ref{fig:Spectrum}, we numerically computed the density of states for various $\sigma$. In the $\sigma\rightarrow 0$ limit, the spectrum has sharp $L=2^{N/2}$ delta-function-like peaks as expected.

\begin{figure}[t!]
 \begin{center}
\includegraphics[scale=0.7]{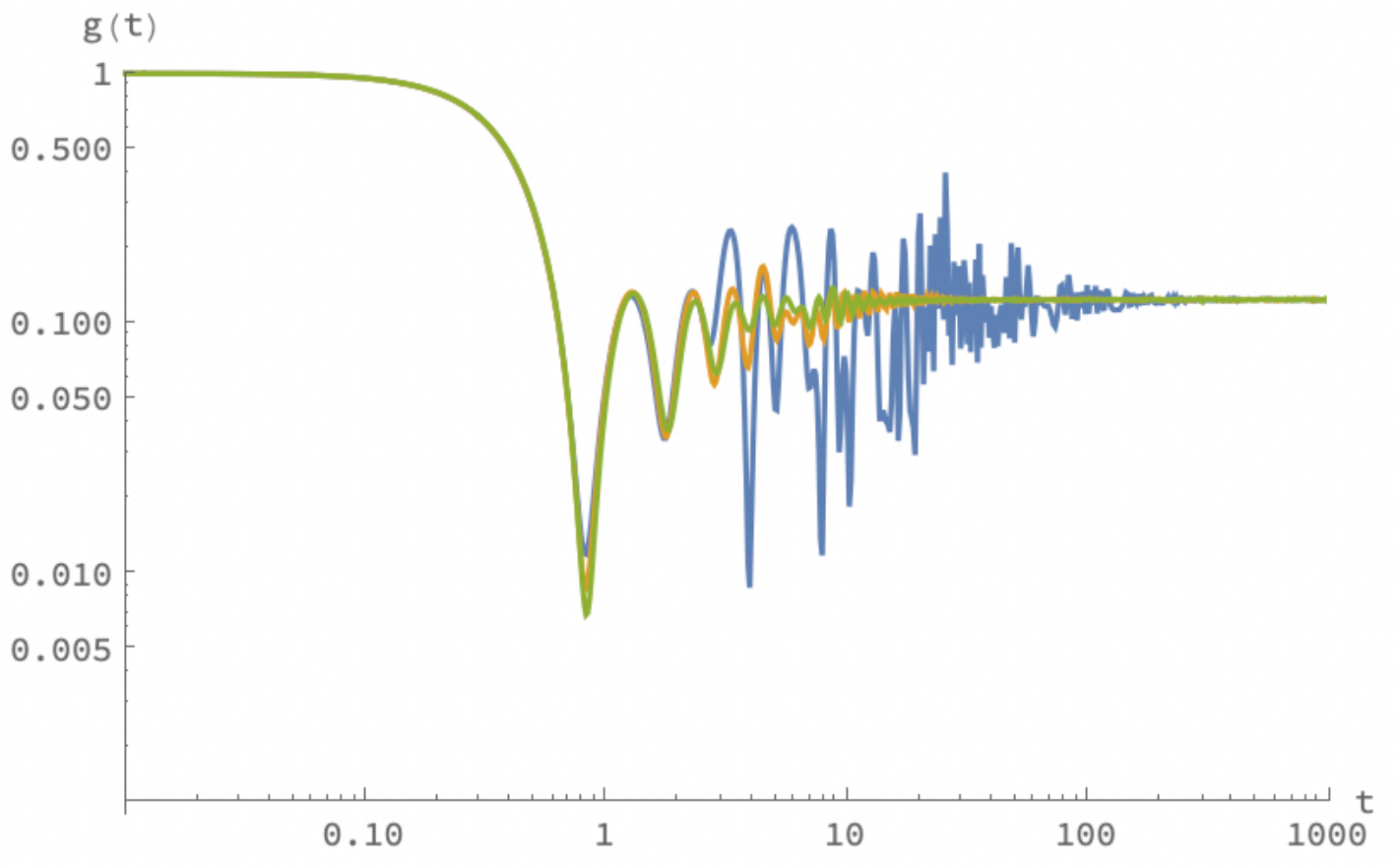}
\end{center}
 \caption{A log-log plot of the spectral form factor $g(t)$ with inverse temperature $\beta=0.0001$, and   $\sigma=0.03$(blue), $0.3$(orange), $3$(green) in the partially disorder-averaged SYK model. We take $N=8$, $q=4$ and $J=1$, and take $50000$ samples for averaging. We have chosen the same value of  the external coupling as Figure~\ref{fig:Spectrum}.}
\label{fig:SFF}
\end{figure}

We also computed  the $\sigma$-dependence of the spectrum form factor 
\begin{align}
    g(t)=\frac{	\big\la Z(\beta+it) Z(\beta-it) \big\ra_J }{\big\la Z(\beta)^2  \big\ra_J}
\end{align}
where 
\begin{align}
    Z(\beta+it)\equiv {\rm Tr}\left(e^{-\beta H-iH t}\right)\, ,
\end{align}
 which probes the discreteness of the spectrum as depicted in Figure~\ref{fig:SFF}. 
We can observe erratic oscillations at late times for small values of $\sigma$, which is characteristic of the SYK model with a fixed coupling \cite{Cotler:2016fpe}.

\section{Conclusions and Discussions}
\label{sec:conclusions}
In this paper, we studied a partially disorder-averaged SYK model.
We introduced a partially disorder-averaged SYK model by modifying the probability distribution of the coupling constants with a real parameter $\s$.
The probability distribution becomes the usual Gaussian form in the $\s \to \inf$ limit
while in the $\s \to 0$ limit, it becomes a product of delta functions which enforces each component of the coupling constant to a fixed external value.
Given this partial disorder averaging, we studied the one- and two-point functions of the partition functions as well as spectral density and the spectral form factor.
For the large $N$ effective description, in addition to the usual bi-local collective fields, we introduced a new additional set of local collective fields.
We explained that these local fields can be understood as the ``half'' of the bi-local collective fields.

\begin{figure}
 \begin{center}
\includegraphics[scale=0.3]{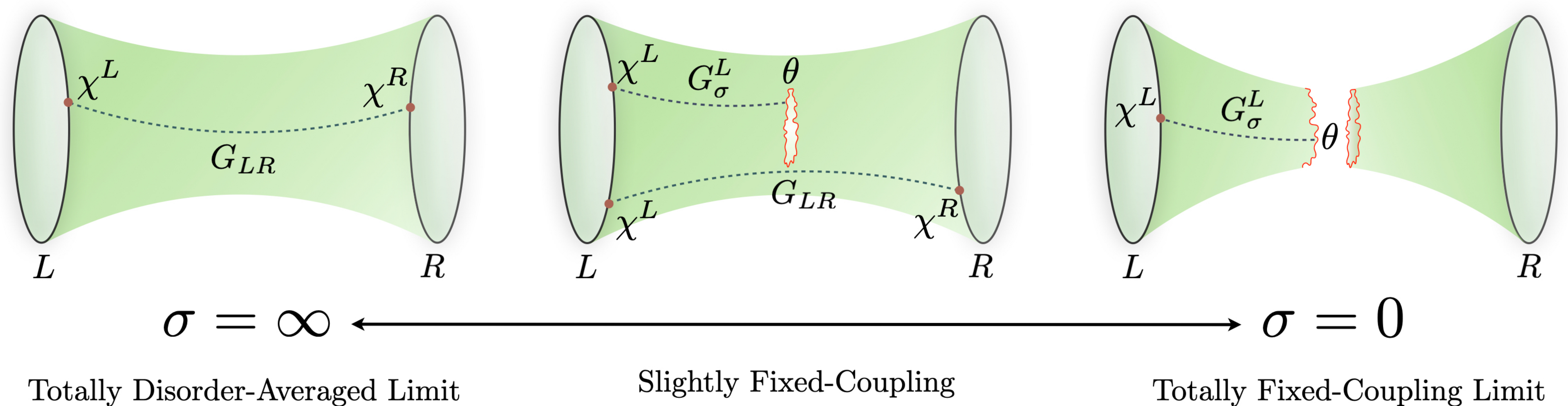}
\end{center}
 \caption{The intuitive bulk pictures dual to the half-wormholes for various values of $\sigma$  in the partially fixed-coupling SYK model. The left picture corresponds to the totally disorder-averaged limit $\sigma=\infty$, where the half-wormhole and the wormhole saddles degenerate. The middle corresponds to the slightly fixed-coupling parameter region, where the half-wormhole saddles can be considered as a small perturbation of the wormhole saddle. The right picture represents the half-wormhole in the totally fixed-coupling limit, where the correlation between the left and the right boundaries totally vanishes.}
\label{fig:HW}
\end{figure}

\bigskip 
{\bf Bulk dual to the half-wormhole saddles}\\
For the study of the two-point function of the zero-dimensional model partition function,
we found that it contains the wormhole and half-wormholes as large $N$ saddles, and their configurations change as we gradually vary $\sigma$. 
A natural question is what is the gravity dual to the half-wormhole for each value of $\sigma$.
As we saw in section \ref{sec:0d SYK2}, in the totally disorder-averaged limit $\sigma=\infty$, the half-wormhole and the wormhole saddles degenerate,
so in this sense, the dual geometry is just a smooth wormhole geometry drawn as the left picture in Figure~\ref{fig:HW}.
Such geometry can be found in the one-dimensional model \cite{Saad:2018bqo} as well as in the dual JT gravity set-up for example in \cite{Saad:2019lba}.

Let us consider the half-wormhole saddles in the slightly fixed coupling region. As we explained section \ref{sec:0d SYK2}, in this region, the half-wormhole  saddle can be obtained by considering a perturbation around the wormhole saddles by adding the small effect of the half-bi-local fields. The interpretation of this perturbation in the gravity side is considered in section \ref{sec:bulk interpretation}. We found that adding the small effect of the half-bi-local fields $G_{\sigma}$ corresponds to having a nontrivial classical configuration of the bulk field dual to $G_\sigma$ on the original wormhole background. As we also pointed out in Appendix \ref{app:external coupling}, the $G_\sigma$ can be interpreted as the correlation between the fermions $\chi_i$ on the boundary and other fermioninc degrees of freedom corresponding to $\theta_i$, which represents the fixed couplings, living on ``another universe". We interpret the gravity dual of ``another universe" as a bulk brane, whose structure strongly depends on the microscopic details of the bulk gravity theory. Such a ``UV-sensitive brane" might appear in the low-energy effective gravity from the condensation of some stringy modes in string theory as pointed out in \cite{Blommaert:2021fob}. We leave it as an interesting research direction for the future.

Similar proposals are also mentioned in \cite{Blommaert:2021gha,Garcia-Garcia:2021squ}. The intuitive picture can be drawn as the middle picture of Figure~\ref{fig:HW}. In this picture, the spacetime ends at the bulk brane placed in the middle of the background wormhole geometry. $G_\sigma$ represents the correlation between the spacetime boundary and this brane. Since we have a nontrivial profile for the bi-local field $G_{LR}$ in this parameter region, the left and right boundaries are still connected by the wormhole. We imagine that as we gradually lower the value of $\sigma$,  the ``gash'' made by the brane is getting bigger, and in the totally fixed-coupling limit $\sigma=0$, the wormhole will be eventually  torn into two pieces (right picture in Figure~\ref{fig:HW}). This is consistent with the fact that the half-wormhole saddle in this limit represents no correlation between the left and right boundaries, i.e., $G_{LR}=0$ while there is a nontrivial profile for the half-bi-local fields $G_{\sigma}$ and $\Sigma_\sigma$. This is also consistent with the bulk picture of the half-wormhole in \cite{Saad:2021rcu}.

\bigskip 
{\bf Linked and unlinked half-wormholes}\\
We analyzed the saddle point solutions for $\big\la Z_LZ_R \big\ra_J$ for large and small values of $\sigma$. We found that there exist both wormhole and half-wormhole saddles except for the totally disorder-averaged limit $\sigma=0$, where these two types of the saddles degenerate. In the totally disorder-averaged limit, only the half-wormhole saddles contribute to $\big\la Z_LZ_R \big\ra^{\sigma=0}_J$, and we can write it schematically as
\begin{align}
    \big\la Z_LZ_R \big\ra^{\sigma=0}_J \approx e^{-S_{\rm on\mathchar`-shell}^{(L)}({\rm Half\mathchar`-wormhole })} e^{-S_{\rm on\mathchar`-shell}^{(R)}({\rm Half\mathchar`-wormhole })}\, .
\end{align}
On the other hand, in \cite{Saad:2021rcu} they analyzed the saddle point solutions in the (totally) fixed-coupling SYK model and found that
\begin{align}
    Z_LZ_R\underset{\rm SSSY}{\approx} e^{-S_{\rm on\mathchar`-shell}^{(LR)}({\rm Wormhole })}+ e^{-S_{\rm on\mathchar`-shell}^{(LR)}({\rm (Linked)\,  Half\mathchar`-wormhole })}\, ,
\end{align}
i.e., the wormhole saddles and the half-wormhole saddles contribute in the same order to $ Z_LZ_R$. Here ``linked half-wormhole" means the half-wormhole saddle in the collective field description where both wormhole and half-wormhole saddle exist. On the other hand, in \cite{Saad:2021rcu} they introduced another collective field description without wormholes.  In this description, the half-wormholes are called ``(unlinked) half-wormholes" to distinguish them from the linked half-wormholes. The half-wormhole saddles expressed as the nontrivial profile of the half-bi-local fields, i.e., $G_\sigma, \Sigma_\sigma\neq 0$ discussed in this paper, would correspond to their unlinked half-wormhole saddles. In this paper, when writing the effective action for the collective fields (\ref{S_eff-0d}) in $0$d SYK, we rescaled them to have nontrivial profiles of the wormhole solutions even in the strictly totally fixed coupling limit. It fits more into the situation of \cite{Saad:2021rcu} given the collective fields without rescaling (\ref{S_eff2}). As you can see, since $J_\sigma\rightarrow J_0$ and $\tilde{J}\rightarrow 0$ in the totally fixed-coupling limit, we can simply integrate over the bi-local field $G_{LR}$, forcing us to $\S_{LR}(\t_1, \t_2) = \frac{1}{2} \S_\s^L(\t_1) \S_\s^R(\t_2)$. As a result, the effective action can be written solely with half-bi-local fields (as well as  bi-local fields of the type $G_{LL},G_{RR},\S_{LL},\S_{RR}$). In this collective field description, it is obvious that there are only half-wormholes described by the half-bi-local fields and no wormholes represented by the solutions $G_{LR}\neq 0, \S_{LR}\neq 0$ with $G_\sigma^{L,R}=\S_\sigma^{L,R}=0$. This situation corresponds to the collective field description with no wormhole saddles discussed in \cite{Saad:2021rcu}.

\bigskip 
{\bf Future directions}\\
In this paper, we studied a partially disorder-averaged Majorana SYK model. Many generalizations of the SYK model were studied by various authors.
Some examples include the complex SYK model \cite{Sachdev:2015efa, Davison:2016ngz, Gu:2019jub}, supersymmetric SYK model \cite{Fu:2016vas, Peng:2017spg, Murugan:2017eto},
SYK model with global symmetries \cite{Gross:2016kjj, Yoon:2017nig},
and higher dimensional generalization of the SYK model \cite{Gu:2016oyy, Berkooz:2016cvq, Turiaci:2017zwd, Liu:2018jhs}. 
It would be interesting to study a partially disorder-averaged version of these generalizations of the SYK model.

The initial motivation to introduce disorder in the original Sachdev-Ye model \cite{Sachdev:1992fk, georges2000mean, georges2001quantum}
was to simulate zero-temperature quantum phase transition between the quantum disordered spin-liquid phase and the magnetically ordered spin-glass phase.
It would be interesting to study whether our partially disorder-averaged model brings any further insights into this aspect.

The original motivation to consider the half-wormholes was to resolve the tension between the factorization of the partition functions and the existence of the wormholes in the bulk in the AdS/CFT context. We found that the wormhole solutions persist even in the fixed coupling limit, but they have zero measure in the path integral.
Instead, the half-wormhole solutions emerge as the ``dissolved''-wormholes, and they dominate the partition function. It would be interesting to study whether a similar thing could happen in the gravitational path integral.

\section*{Acknowledgements}
We are grateful to Tomoki Nosaka for sharing the Mathematica code for computing the spectral form factor in the SYK model.
We also thank Takanori Anegawa, Norihiro Iizuka, and Tadashi Takayanagi for useful discussion.
KG is supported by JSPS Grant-in-Aid for Early-Career Scientists 21K13930.
The work of KS is supported by the Simons Foundation through the ``It from Qubit'' collaboration. 
TU was supported by JSPS Grant-in-Aid for Young Scientists  19K14716 and MEXT KAKENHI Grant-in-Aid for Transformative Research Areas A “Extreme Universe” No.21H05184.
This work is also supported by MEXT KAKENHI Grant Number 21H05182, 21H05184, and 21H05187.

\appendix
\section{External couplings as another universe fermions}
\label{app:external coupling}
In this appendix, we discuss the interpretation of the $\th_i$ variables introduced in (\ref{J_0}) as non-dynamical fermions living in another universe.
To consider another universe, we start from the two-point function (\ref{Z_LZ_R}).
Taking the total disorder-averaging ($\s=\inf$) of this function, we find
	\begin{align}
		\big\la Z_L Z_R \big\ra_J \, = \, \int D\c^L_i D\c^R_i \, &\exp\Bigg[-\frac{1}{2} \sum_{a=L,R} \int d\t \sum_{i=1}^N \c^a_i \pa_\t \c^a_i \nn\\
		&\ \qquad + \, \frac{(q-1)!\wtd{J}^2}{2N^{q-1}} \sum_{i_1< \cdots <i_q}^N \left( \int d\t \sum_{a=L,R} \c^a_{i_1} \cdots \c^a_{i_q} \right)^2 \Bigg] \, .
 	\end{align}
Now we freeze the dynamics of $\c^R_i$ and denote the non-dynamical fermions as $\th_i \propto \c^R_i$.
Therefore, now we can interpret the above two-point function as a one-point function of $Z_L$ with external fermionic sources $\th_i$:
	\begin{align}
		\big\la Z_L \big\ra_J \, = \, \int D\c_i \, &\exp\Bigg[-\frac{1}{2} \int d\t \sum_{i=1}^N \c_i \pa_\t \c_i
		\, + \, \frac{J_\s}{qN^{q-1}} \int d\t \left( \sum_{i=1}^N \th_i \c_i(\t) \right)^q\nn\\
		&\qquad \qquad + \, \frac{\wtd{J}^2}{2qN^{q-1}} \int d\t_1 d\t_2 \left( \sum_{i=1}^N \c_i(\t_1) \c_i(\t_2) \right)^q \Bigg] \, ,
 	\end{align}
where we denoted $\c_i \equiv \c^L_i$ and also adjusted the proportionality constant of $\th_i \propto \c^R_i$ to give the precise second term.
This effective action precisely agrees with (\ref{seff}), and in this sense, we can interpret the external variable $\th_i$ as non-dynamical fermions living in another universe.

\section{Integration over the fermions}
\label{app:integration}
In this appendix, we present some details for the integration over fermions.
This is different from the usual Gaussian integrals of Grassmann variables due to the external variables $\th_i$ (\ref{J_0}) and its anti-commutation relation
	\begin{align}
		\{\th_i, \th_j\} \, = \, \d_{ij} \, .
  	\end{align}
To implement this anti-commutation relation in the path-integral,
we can regard this $\th_i$ as $N$-dimensional Dirac matrices constructed from the Pauli matrix (for example see \cite{Sarosi:2017ykf}).  

With this representation of $\th_i$,
we first consider the fermion integration in the one-point function (of the one-dimensional model) studied in section~\ref{sec:partition function}:
	\begin{align}
		\int \prod_{i=1}^N D\c_i \, \exp \Bigg[ \frac{1}{2} \int d\t_1 d\t_2 \sum_{i=1}^N \c_i(\t_1) \S(\t_1, \t_2) \c_i(\t_2) 
	    \, + \int d\t \sum_{i=1}^N \c_i(\t) \S_\s(\t) \th_i \Bigg] \, .
  	\end{align}
For the kinetic term, using the delta function $\d(G - N^{-1} \sum_i \c_i \c_i)$, we rewrite it in terms of the bi-local field $G(\t_1,\t_2)$,
so that it does not appear in the fermion integral.
As usual Majorana fermion integrals, we introduce complex $N/2$ fermions $c_i$ and also complex $N/2$ external variables $\vp_i$ by 
	\begin{gather}
		\c_{2i} \, = \, \frac{c_i+\bar{c}_i}{\sqrt{2}} \, , \qquad \quad \c_{2i-1} \, = \, \frac{i(c_i-\bar{c}_i)}{\sqrt{2}} \, , \\
		\th_{2i} \, = \, \frac{\vp_i+\bar{\vp}_i}{\sqrt{2}} \, , \qquad \quad \th_{2i-1} \, = \, \frac{i(\vp_i-\bar{\vp}_i)}{\sqrt{2}} \, .
  	\end{gather}
Then, the above integral is given by
	\begin{align}
		&\int \prod_{i=1}^{N/2} Dc_i D\bar{c}_i \, \exp \Bigg[ \int d\t_1 d\t_2 \sum_{i=1}^{N/2} c_i(\t_1) \S(\t_1, \t_2) \bar{c}_i(\t_2) 
	    \, + \int d\t \sum_{i=1}^{N/2} \S_\s (\vp_i \bar{c}_i + \bar{\vp}_i c_i ) \Bigg] \nn\\
	    &= \, \Big[ \Pf\left( \S + \tfrac{1}{2} \S_\s \S_\s \right) \Big]^N \, .
  	\end{align}
Exponentiating this into the effective action and sifting $\S(\t_1, \t_2) \to \S(\t_1, \t_2) - \frac{1}{2} \S_\s(\t_1) \S_\s(\t_2)$, we find
	\begin{align}
		&S_{\rm eff}[G, \S, G_\s, \S_\s] \, = \, - \frac{N}{2}\Tr\log(\S ) - \frac{N}{2} \int d\t \big[ \pa_\t G(\t, \t')\big]_{\t'=\t} \\
		&\qquad \qquad \qquad  +\frac{N}{2} \int d\t_1 d\t_2 \left( \S(\t_1, \t_2) G(\t_1, \t_2) - \frac{\wtd{J}^2}{q} G(\t_1, \t_2)^q \right) \nn\\
		&\qquad + N \int d\t \left( \S_\s(\t) G_\s(\t) - \frac{J_\s}{q} G_\s(\t)^q \right)
		\, - \, \frac{N}{4} \int d\t_1 d\t_2 \, \S_\s(\t_1) G(\t_1, \t_2) \S_\s(\t_2) \, . \nn
	\end{align}

For the zero-dimensional case in section~\ref{sec:0d SYK1}, we do not have the bi-local fields. Therefore, the fermion integral is simply 
	\begin{align}
		\int \prod_{i=1}^N d\c_i \, \exp\bigg[ - \S_\s \sum_{i=1}^N \th_i \c_i \bigg] \, = \, \big( -\S_\s \big)^N \bigg( \prod_{i=1}^N \th_i \bigg) \, .
  	\end{align}

Finally, let us consider the fermion integration for the two-point function in the zero-dimensional SYK model studied in section~\ref{sec:0d SYK2}.
Since we have 
	\begin{align}
		&\quad \int d\c^L d\c^R \, \exp\Big[ \S_{LR} \c^L \c^R\Big]\exp\Big[ \S_\s^L \th \c^L\Big]\exp\Big[ \S_\s^R \th \c^R \Big] \nn\\
		&= \, \int d\c^L d\c^R \Big( 1 + \S_{LR} \c^L \c^R + \th^2 \S_\s^L \S_\s^R \c^L \c^R + \cdots \Big) \nn\\
		&= \, \S_{LR} + \frac{1}{2} \, \S_\s^L \S_\s^R \, ,
  	\end{align}
where we used $\th^2 = 1/2$, the integrating out $\c^L_i \c^R_i$ gives $-N\log(\S_{LR} + \frac{1}{2} \S_\s^L \S_\s^R)$ in the effective action.
Sifting $\S_{LR} \to \S_{LR}- \frac{1}{2} \S_\s^L \S_\s^R$, we find
	\begin{align}
		S_{\rm eff}[G_{LR}, \S_{LR}, G_\s^a, \S_\s^a] \, &= \, - N \log (\S_{LR}) \, + \, N \left( \S_{LR} G_{LR} - \frac{\wtd{J}^2}{q} \, G_{LR}^q \right) \nn\\
		&\quad +N \sum_a \left( \S_\s^a G_\s^a - \frac{J_\s}{q} \big( G_\s^a \big)^q \right) \, - \, \frac{N}{2} G_{LR} \S_\s^L \S_\s^R\, .
	\end{align}
To be precise, we should regard $G_\s^a$ as a matrix as
	\begin{align}
        \big[ G_\sigma^L, G_\sigma^R \big] \, &= \, \frac{1}{N^2}\sum_{i,j} \big[ \theta_i \chi_i^L, \theta_j \chi_j^R \big] \nn\\
        &= \, \frac{1}{N^2} \sum_{i,j} (\theta_i \theta_j + \theta_j \theta_i) \chi_i^L \chi_j^R \nn\\
        &= \, \frac{1}{N^2} \sum_i \chi_i^L \chi_i^R \nn\\
        &= \, \frac{1}{N} \, G_{LR} \, .
	\end{align}
However, as long as we keep the large $N$ limit, we can treat $G_\s^a$ as a usual $c$-number.

\section{Another formalism for slightly fixed-coupling}
\label{app:schwarzian}
In this appendix, we will show that the same structure we saw in section \ref{sec:slightly fixed-coupling} and \ref{sec:bulk interpretation}
also appears in another formulation of the slightly fixed-coupling SYK model.
For simplicity, in this appendix, we focus on the $q=4$ case.

For this purpose, instead of (\ref{HS-trick}), we introduce another Hubbard–Stratonovich trick
	\begin{align}
		1 \, &= \, \int DG \int DG^\s_{ij} \, \d\left( G(\t_1, \t_2) - \frac{1}{N} \sum_{i=1}^N \c_i(\t_1) \c_i(\t_2) \right) \d\Big( G^\s_{ij}(\t) - \c_i(\t) \c_j(\t) \Big) \nn\\
		&= \, \int DG D\S \int DG^\s_{ij} D\S^\s_{ij} \exp\Bigg[ - \frac{1}{2} \int d\t_1 d\t_2 \, \S(\t_1, \t_2) \left( N G(\t_1, \t_2) - \sum_{i=1}^N \c_i(\t_1) \c_i(\t_2) \right) \nn\\
		&\hspace{160pt}  - \frac{1}{2} \sum_{i,j=1}^N \int d\t \, \S^\s_{ij}(\t) \Big( G^\s_{ij}(\t) - \c_i(\t) \c_j(\t) \Big) \Bigg] \, .
 	\end{align}
Using this trick for the partially disorder-averaged partition function and performing the Gaussian integral for $\c_i$, we obtain
	\begin{align}
		\big\la Z \big\ra_J \, = \, \int DG D\S DG^\s D\S^\s \, e^{-S_{\rm eff}[G, \S, G^\s, \S^\s]} \, ,
 	\end{align}
with
	\begin{align}
		S_{\rm eff}[G, \S, G^\s, \S^\s] \, &= \, - \frac{1}{2} \Tr \log (\D_{ij}) + \frac{N}{2} \int d\t_1 d\t_2 \left( \S(\t_1, \t_2) G(\t_1, \t_2) - \frac{\wtd{J}^2}{4} G(\t_1, \t_2)^4 \right) \nn\\
		&\qquad + \frac{1}{2} \sum_{i,j=1}^N \int d\t \left( \S^\s_{ij}(\t) G^\s_{ij}(\t) + \frac{\wtd{J}^2}{12\s^2} \sum_{k,l=1}^N J_{ijkl}^{(0)} G^\s_{ij}(\t) G^\s_{kl}(\t) \right) \, ,
	\end{align}
where the trace is now taken both for the indices $i,j$ and the bi-local time $(\t_1, \t_2)$ with
	\begin{align}
		\D_{ij}(\t_1, \t_2) \, \equiv \, \d_{ij} \Big( \d(\t_1 - \t_2) \pa_{\t_2} - \S(\t_1, \t_2) \Big) - \S^\s_{ij}(\t_1) \d(\t_1 - \t_2) \, .
 	\label{Delta_ij}
	\end{align}
If this $\D_{ij}$ is completely diagonal for the indices $i,j$, we recover a factor of $N$ in front of the trace term, but for a general $\D_{ij}$, we miss this factor of $N$.
The extra local fields $\{G^\s, \S^\s\}$ represents a non-singlet sector of the global $O(N)$ symmetry.
In general, it is hard to understand the non-singlet sector of a large $N$ theory \cite{Gross:1990md, Maldacena:2005hi}.
In the context of AdS$_2$/CFT$_1$, these fields are probably related to the open string degrees of freedom attached to the boundary (see section 6.2 of \cite{Maldacena:2016hyu}).

Even though we miss the factor of $N$ in front of the trace term and the second line of the above action, let us here pretend that we can use the large $N$ saddle-point evaluation.
Variation of each field leads to saddle-point equations
	\begin{align}
		\d G \, :& \qquad \S(\t_1, \t_2) \, = \, \wtd{J}^2 \, G(\t_1, \t_2)^3 \, , \\
		\d \S \, :& \qquad \frac{1}{N} \sum_{i=1}^N \Big[\D^{-1}(\t_1, \t_2) \Big]_{ii} \, = \, G(\t_1, \t_2) \, , \label{delta Sigma-app} \\
		\d G^\s \, :& \qquad \S^\s_{ij}(\t) \, = \, - \, \frac{\wtd{J}^2}{6\s^2} \, \sum_{k,l=1}^N \, J_{ijkl}^{(0)} G^\s_{kl}(\t) \, , \label{delta G_sigma-app} \\
		\d \S^\s \, :& \qquad \Big[ \D_{ij}^{-1} \Big](\t, \t) \, = \, G^\s_{ij}(\t) \, , 
  	\end{align}
where we defined two kinds of inverse. The first one is inverse for the bi-local time:
	\begin{align}
		\int d\t_3 \, \D_{ij}(\t_1, \t_3) \Big[ \D^{-1}(\t_3, \t_2) \Big]_{ij} \, = \, \d(\t_1 - \t_2) \, , \qquad  \quad ({\rm for\ fixed} \ i, \, j) \, , 
 	\end{align}
and the second one is for the $O(N)$ index:
	\begin{align}
		\sum_{k=1}^N \, \D_{ik}(\t_1, \t_2) \Big[ \D^{-1}_{kj} \Big](\t_1, \t_2) \, = \, \d_{ij} \, , \qquad  \quad ({\rm for\ fixed} \ \t_1, \, \t_2) \, .
 	\end{align}

We first note that the saddle-point solution of the diagonal part of $G^\s$ must be given by
	\begin{align}
		G^\s_{ii}(\t) \, = \, \big\la \c_i(\t) \c_i(\t) \big\ra \, = \, \frac{1}{2} \, ,
  	\label{G_ii}
	\end{align}
where for the second equality, we used the equal-time anti-commutation relation of the fermion (\ref{ACR}).
Let us now consider the following external coupling $J_{ijkl}^{(0)}$
	\begin{align}
		\qquad J_{ijkl}^{(0)} \, = \, \frac{J_0}{6} \Big( \d_{ij} \d_{ik} \d_{il} \, + \, A_{ijkl}^{(0)} \Big) \, ,
	\end{align}
where $A_{ijkl}^{(0)}$ is the totally anti-symmetric tensor without any diagonal piece. Namely, if any of the indices coincide, $A_{ijkl}^{(0)}=0$.
Then, from (\ref{delta G_sigma-app}) the index structure of $\S^\s_{ij}$ is decomposed into symmetric and antisymmetric parts as
	\begin{align}
		\S^\s_{ij}(\t) \, = \, \d_{ij} \S^\s_{\rm dig}(\t) \, + \, \S^{\rm ant}_{ij}(\t) \, ,
  	\end{align}
and the equation (\ref{delta G_sigma-app}) is reduce to 
	\begin{align}
		\S^\s_{\rm dig}(\t) \, &= \, - \, \frac{J_\s}{N} \, \sum_{k=1}^N \, G^\s_{kk}(\t) \, = \, - \frac{J_\s}{2}\, , \\
		\S^{\rm ant}_{ij}(\t) \, &= \, - \, \frac{J_\s}{N} \, \sum_{k,l=1}^N \, A_{ijkl}^{(0)} G^\s_{kl}(\t) \, .
  	\label{Siga^sigma eq}
	\end{align}
Using this expression of $\S^\s_{ij}$ into (\ref{Delta_ij}), we can see that the diagonal components of $\D_{ij}$ are independent of the index $i$
	\begin{align}
		\D_{ii}(\t_1, \t_2) \, &= \, \d(\t_1 - \t_2) \left( \pa_{\t_2} + \frac{J_\s}{2} \right) \, - \, \S(\t_1, \t_2) \nn\\
		&\equiv \, \D_\Dig(\t_1, \t_2) \, .
	\label{Delta_Dig}
	\end{align}
Therefore, the summation in (\ref{delta Sigma-app}) is trivially taken and gives a factor of $N$, which cancels with the $1/N$ coefficient.
Hence, we can simply invert $\D^{-1}(\t_1, \t_2)$ and this equation is now written as
	\begin{align}
		\d(\t_1 - \t_2) \, = \, \int d\t_3 \, \D_\Dig(\t_1, \t_3) G(\t_3, \t_2) \, .
  	\end{align}
The modification from the ordinary SYK is the $J_\s$ term in (\ref{Delta_Dig}), which acts as a mass or chemical potential term.

\section{Two-point function of $q=2$, 0d SYK}
\label{app:q=2}
In this appendix, we study the $q=2$ case of the $0d$ SYK model discussed in section~\ref{sec:0d SYK2}.
For $q=2$, we have the saddle-point equations
	\begin{gather}
		\S_{LR} \, = \, G_{LR} \, + \, \l_\s \, \S_\s^L \S_\s^R \, , \qquad 
		\frac{1}{\S_{LR}} \, = \, G_{LR} \, , \label{G^LR-eq-app} \\[2pt]
		\S_\s^a \, = \, G_\s^a \, , \qquad
		G_\s^L \, = \, \l_\s G_{LR} \S_\s^R \, , \qquad 
		G_\s^R \, = \, \l_\s G_{LR} \S_\s^L \, . \label{Sigma_sigma-eq-app}
  	\end{gather}
From (\ref{Sigma_sigma-eq-app}), we find equations for $G_\s^L$ as
	\begin{align}
	    \Big[ \big( \l_\s G_{LR} \big)^2 + 1 \Big] G_\s^L \, = \, 0 \, .
	\end{align}
Therefore, the trivial solution is $\{ G_\s^L \, , G_\s^R \} = \{0, 0\}$, which can be understood as the ``wormhole'' saddle-point solution again.
The nontrivial solution can be found only if 
	\begin{align}
		G_{LR} \, = \, \pm \, \l_\s^{-1} \, .
	\end{align}
In this case, $\{ G_\s^L \, , G_\s^R \}$ are not yet fixed by the equations (\ref{Sigma_sigma-eq-app}).
Since in the totally fixed-coupling limit ($\s \to 0$), we have $\l_\s^{-1} \propto \s^{2/q}$, this solution behaves as $G_{LR} \to 0$ in this limit.
Therefore, this nontrivial solution is understood as the ``half-wormhole'' saddle-point solution.

For the wormhole case, the solution of the bi-local sector is again given by (\ref{wormhole}) as $\{G_{LR} , \S_{LR}\} = \{\pm 1, \pm 1\}$.
For the half-wormhole case, solving (\ref{G^LR-eq-app}), we find 
	\begin{align}
		\{ G_\s^L \, , G_\s^R \} \, = \, \left\{ \pm\sqrt{1-\l_\s^{-2}} \, , \, \sqrt{1-\l_\s^{-2}} \right\} \, .
  	\end{align}

The on-shell actions are obtained 
	\begin{align}
		S^{\onshell}_{\WH} \, = \, \frac{N}{2} \, + \, m\pi i N \, ,
	\end{align}
for the wormhole saddles with $m=1,2$, and 	
	\begin{align}
		S^{\onshell}_{\HW} \, \approx \, - \, N \log( \l_\s ) \, + \, N \sqrt{1-\l_\s^{-2}} \, + \, n\pi i N \, .
	\end{align}
for the half-wormhole solutions with $n=1,2$.

\section{Direct computation of $\big\la Z_L Z_R \big\ra_J$ at $\s=0$}
\label{app:direct computation}
In this appendix, we directly solve the two-point function (\ref{<Z_LZ_R>}) at $\s=0$.
Because the prefactor $\wtd{J}^{2N/q} \to \s^{2N/q}$ in the $\s\to 0$ limit, while the coupling in the effective action (\ref{S_eff-0d}) is $\l_\s \to \s^{-2/q}$ in this limit,
so that the action gives a divergence.
To get a finite non-zero result,
we have to expand the interaction term of (\ref{S_eff-0d}) perturbatively and consider the $N$-th order term (which gives $\l_\s^N$).
Therefore, the two-point function at $\s=0$ is given by
	\begin{align}
		\big\la Z_L Z_R \big\ra_J^{\s=0} \, &= \, \frac{N^N J_0^{\frac{2N}{q}}}{2^N N!}
		\int dG_{LR} \, \frac{d\S_{LR}}{2\pi i/N} \int \prod_a^{L,R} dG_\s^a \, \frac{d\S_\s^a}{2\pi i/N} \, \Big( G_{LR} \S_{LR} \S_\s^L \S_\s^R \Big)^N \nn\\
		&\quad \times \exp\left[ -N \left( \S_{LR} G_{LR} - \frac{1}{q} G_{LR}^q \right) -N \sum_a \left( \S_\s^a G_\s^a - \frac{1}{q} (G_\s^a)^q \right) \right] \, .
 	\end{align}
Now we can see that each set of $\{G, \S\}$ decoupled from each other.
Therefore, we can perform the integrals separately. For the bi-local sector, we have
	\begin{align}
		& \int dG_{LR} \, \frac{d\S_{LR}}{2\pi i/N} \, \Big( G_{LR} \S_{LR} \Big)^N \exp\left[ -N \left( \S_{LR} G_{LR} - \frac{1}{q} G_{LR}^q \right) \right] \nn\\
		= \, &\int dG_{LR} \, e^{\frac{N}{q} G_{LR}^q} \big( G_{LR} \big)^N \big( - N^{-1} \pa_{G_{LR}} \big)^N \d(G_{LR}) \nn\\
		= \, & \, \frac{N!}{N^N} \, ,
 	\end{align}
while for the local sector, we have
	\begin{align}
		& \int dG_\s^L \, \frac{d\S_\s^L}{2\pi i/N} \, \big( \S_\s^L \big)^N \exp\left[ -N \left( \S_\s^L G_\s^L - \frac{1}{q} (G_\s^L)^q \right) \right] \nn\\
		= \, &\int dG_\s^L \, e^{\frac{N}{q} (G_\s^L)^q} \big( - N^{-1} \pa_{G_\s^L} \big)^N \d(G_\s^L) \nn\\
		= \, & \, \frac{N!}{N^N} \frac{(N/q)^{N/q}}{(N/q)!}\, .
 	\end{align}
Therefore, combining these integral results, we find that the two-point function at $\s=0$ is given by
	\begin{align}
		\big\la Z_L Z_R \big\ra_J^{\s=0} \, &= \, \frac{J_0^{\frac{2N}{q}}}{2^N} \left( \frac{N!}{N^N} \frac{(N/q)^{N/q}}{(N/q)!} \right)^2 \nn\\
		&\approx \, 2^{-N} J_0^{\frac{2N}{q}} \, q \, e^{- 2N\left( 1 - \frac{1}{q} \right)} \, ,
 	\end{align}
where for the last line, we used the large $N$ approximation besides the factor $2^{-N} J_0^{2N/q}$.
This result agrees with the saddle-point evaluation (\ref{<Z_LZ_R>-small-sigma}).


\bibliographystyle{JHEP}
\bibliography{Refs} 



\end{document}